\documentclass[12pt]{iopart}
\usepackage{amstext,iopams,amsfonts,amssymb,amsbsy,amsmath}
\usepackage{bm}
\usepackage{color,lscape}
\usepackage{graphicx}
\usepackage{multirow}
\usepackage{rotating}
\begin{document}

\def\de{\mathrm{d}}
\def\rmT{\mathrm{T}}
\def\rmBH{\mathrm{BH}}
\def\rmDVCS{\mathrm{DVCS}}
\def\rmI{\mathrm{I}}
\def\rmunp{\mathrm{UU}}
\def\rmTP{\mathrm{UT}}
\def\rmRe{\mathrm{Re}}
\def\rmIm{\mathrm{Im}}
\def\pgate{{\mathcal P}_1(\phi){\mathcal P}_2(\phi)}

\def\CalH{\mathcal{H}}
\def\CalE{\mathcal{E}}
\def\CalHtil{\widetilde{\mathcal{H}}}
\def\CalEtil{\widetilde{\mathcal{E}}}

\def\CalN{\mathcal{N}}
\def\CUU{\sigma_{UU}}
\def\Lumi{\mathcal{L}\,}
\def\xx{\bm{x}}
\def\tt{\bm{\eta}}
\def\ttDVCS{\bm{\eta}_{\mathrm{UT}}^{\rmDVCS}}
\def\ttC{\bm{\eta}_\mathrm{C}}
\def\ttI{\bm{\eta}_{\mathrm{UT}}^{\rmI}}
\def\AC{A_\mathrm{C}}
\def\AUT{A_\mathrm{UT}}
\def\AUTDVCS{A_{\mathrm{UT},\rmDVCS}}
\def\AUTI{A_{\mathrm{UT},\rmI}}
\def\CalAC{\mathcal{A}_\mathrm{C}}
\def\CalAUTDVCS{\mathcal{A}_{\mathrm{UT}}^{\rmDVCS}}
\def\CalAUTI{\mathcal{A}_{\mathrm{UT}}^{\rmI}}

\def\Nobs{N_{\mathrm{o}}}
\def\Nexp{N_{\mathrm{e}}}   % not used
\def\intN{{\mathcal N}}
\def\intNpar{{\mathcal N}_{\mathrm{par}}}
\def\intNpartil{\widetilde{\mathcal N}_{\mathrm{par}}}
\def\calA{{\mathcal A}}
\newcommand{\lbrase}{\left\{}
\newcommand{\rbrase}{\right\}}
\newcommand{\Lbrase}{\Bigl\{}
\newcommand{\Rbrase}{\Bigr\}}

% \begin{frontmatter}

\title[HERMES: DVCS On Transversely Polarised Hydrogen]
{Measurement of Azimuthal Asymmetries With Respect To Both Beam Charge and Transverse Target 
Polarization in Exclusive Electroproduction of Real Photons \\
%  \underline{Paper Tag: dvcsAUT (V6.8)}
%  \hspace*{1cm}\underline{Drafting Committee: 68} \\
\centerline{HERMES Collaboration}
}

% \collab{HERMES Collaboration}

\author{
A.~Airapetian$^{15}$,
N.~Akopov$^{26}$,
Z.~Akopov$^{26}$,
A.~Andrus$^{14}$,
E.C.~Aschenauer$^{6}$,
W.~Augustyniak$^{25}$,
R.~Avakian$^{26}$,
A.~Avetissian$^{26}$,
E.~Avetisyan$^{5,10}$,
L.~Barion$^{9}$
% ...
}
% ...
\address{$^{5}$DESY, 22603 Hamburg, Germany}
\address{$^{6}$DESY, 15738 Zeuthen, Germany}
% ...
\address{$^{9}$Istituto Nazionale di Fisica Nucleare, Sezione di Ferrara and Dipartimento di Fisica, Universit\`a di Ferrara, 44100 Ferrara, Italy}
\address{$^{10}$Istituto Nazionale di Fisica Nucleare, Laboratori Nazionali di Frascati, 00044 Frascati, Italy}
% ...
\address{$^{14}$Department of Physics, University of Illinois, Urbana, Illinois 61801-3080, USA}
\address{$^{15}$Randall Laboratory of Physics, University of Michigan, Ann Arbor, Michigan 48109-1040, USA }
% ...
\address{$^{25}$Andrzej Soltan Institute for Nuclear Studies, 00-689 Warsaw, Poland}
\address{$^{26}$Yerevan Physics Institute, 375036 Yerevan, Armenia}
% ...

\vspace*{5pt}
\ead{management@hermes.desy.de}

\maketitle
%\newpage
\begin{abstract} 
Azimuthal asymmetries in exclusive electroproduction of real photons are measured for the first time with respect to transverse target polarisation, 
providing new constraints on Generalized Parton Distributions. 
From the same data set on a hydrogen target, new results for the beam-charge asymmetry are also extracted with better precision than those previously reported. 
By comparing model calculations with measured asymmetries 
%most importantly to those 
attributed to the interference between the deeply virtual Compton scattering and Bethe-Heitler processes,
%\footnote{Zhenyu: this is not anymore true after including $\AUTDVCS^{\sin{(\phi-\phi_S)}}$ into the game!} 
a model-dependent constraint is obtained on the total angular momenta carried by up and down quarks in the nucleon. 
\end{abstract}

%\begin{keyword}
% keywords, in the form: keyword \sep keyword

% PACS codes, in the form: \PACS code \sep code
%\end{keyword}

% \end{frontmatter}

% -------------------- Main Text --------------------

%\section{}
%\label{}

% \setvruler[15pt][1][1][3][1][30pt][30pt][-42pt]

\section{Introduction}

The partonic structure of the nucleon has traditionally been described in terms of Parton Distribution Functions (PDFs) of the parton's longitudinal momentum as a fraction of the nucleon's momentum in a frame in which the nucleon is 
moving at almost the velocity of light. 
These functions appear in the theoretical description of, 
e.g., Deep-Inelastic Scattering (DIS). 
However, in the context of the rapid theoretical developments of the last decade, PDFs have been conceptually 
subsumed within the broader framework of Generalized Parton Distributions (GPDs), which also describe  elastic form factors and amplitudes for 
hard-exclusive reactions leaving the target nucleon 
intact~\cite{Mue94,Ji97,Rad97}. Most often discussed are the four twist-2 quark-chirality conserving 
quark GPDs: the polarisation-independent distributions $H_q$ and $E_q$ and the polarisation-dependent
distributions $\widetilde{H}_q$ and $\widetilde{E}_q$. The GPDs $H_q$ and $\widetilde{H}_q$ conserve
nucleon helicity, while $E_q$ and $\widetilde{E}_q$ are associated with a helicity
flip of the nucleon.
GPDs depend on the kinematic variables $x$ and $\xi$,
which represent respectively
the average and difference of the longitudinal momentum fractions of the probed parton in the initial and final states. The variable $\xi$ is typically nonzero  in hard-exclusive reactions. GPDs also depend on the squared four-momentum transfer $t=(p-p^\prime)^2$ 
to the nucleon, with $p$ ($p^\prime$) the four-momentum of 
the nucleon in the initial (final) state. 
PDFs and nucleon elastic form factors appear as kinematic limits 
($t \rightarrow 0$) 
and $x$-moments of GPDs, respectively. 
Strong interest in the formalism of GPDs and their experimental constraint has emerged after moments of certain GPDs were found to relate directly to the total (including orbital) angular momenta carried by partons in the nucleon, via the Ji relation~\cite{Ji97}:
\begin{equation}
\lim_{t\to 0} \int_0^1 \de x\ x\left(H_q(x,\xi,t) + E_q(x,\xi,t)\right) = J_q \label{eq:JiRelation}
\end{equation} 
This finding offers for the first time a path towards solving the `nucleon spin puzzle' of how the helicities and orbital angular momenta of quarks and gluons combine to form the spin of the nucleon. 
More recent discussions have focused on the potential of GPDs as multi-dimensional representations of hadrons at the partonic level, correlating the longitudinal momentum fraction with transverse spatial coordinates~\cite {Bur00,Die02,Ral02,Bel02,Bur03}.
\begin{figure}
\begin{center}
\includegraphics[width=0.28\columnwidth,angle=270]{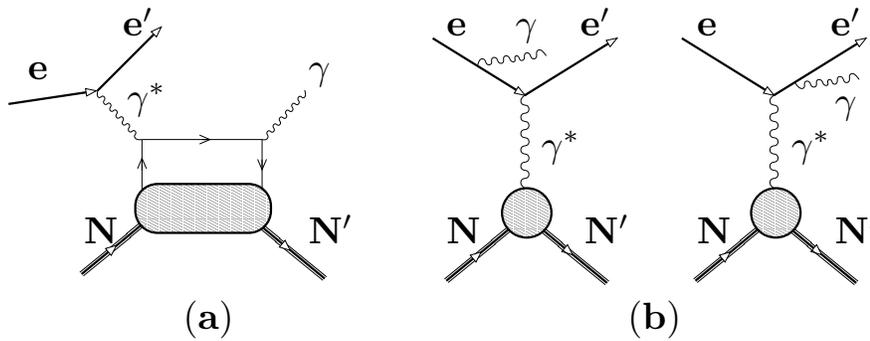}
\end{center}
\caption{Leading-order diagrams for (a) deeply virtual Compton scattering \\ (handbag diagram)  and (b) Bethe-Heitler processes.\label{fig:feymann}}
\end{figure}

\section{Deeply virtual Compton scattering}

\begin{figure} 
\begin{center} 
\includegraphics[width=0.35\columnwidth,angle=270]{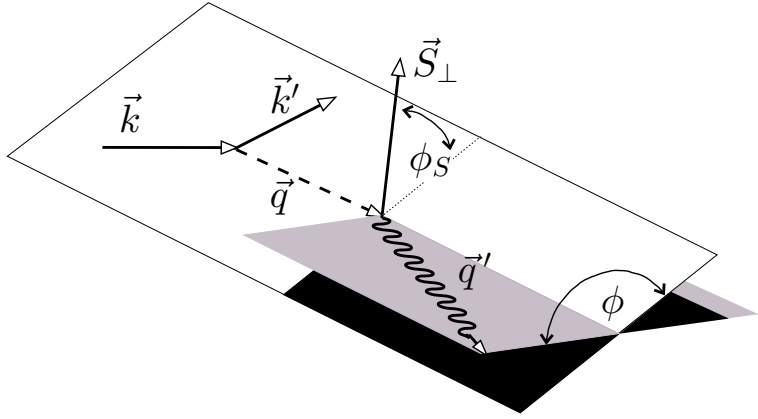} 
\end{center} 
\caption{Momenta and azimuthal angles for exclusive 
electroproduction of photons in the target rest frame.
The quantity $\phi$ denotes the angle between the lepton plane containing the three-momenta $\vec{k}$ and $\vec{k}^\prime$ of the incoming and outgoing lepton and the plane correspondingly defined by 
$\vec{q}=\vec{k}-\vec{k}^\prime$ and the momentum $\vec{q}\,^\prime$ of the real photon. 
The symbol $\phi_S$ denotes the angle between the lepton plane and $\vec{S}_\perp$, the component of the target polarisation vector 
that is 
orthogonal to $\vec{q}$. 
These definitions are consistent with the Trento conventions~\cite{Bac04}.
 \label{fig:coordinate}} 
\end{figure}

Generalized Parton Distributions are accessible through
exclusive processes that involve at least two hard vertices, yet leave the target nucleon intact. 
Among all presently practical hard exclusive probes, the Deeply Virtual Compton Scattering (DVCS) process, i.e., the hard exclusive leptoproduction of a real photon \mbox{(e.g., $\gamma^*\, N \rightarrow \gamma\, N^\prime$)}, appears to have the most reliable interpretation in terms of GPDs.
In electroproduction, direct access to the DVCS amplitude $\rmT_{\rmDVCS}$ is provided by the interference between the DVCS and Bethe-Heitler (BH) processes, in which the photon is radiated from a quark and from the lepton, respectively (see Fig.~\ref{fig:feymann}). 
Since these processes are intrinsically indistinguishable,
%have identical asymptotic intial and final states, 
the cross section is proportional to the squared photon-production amplitude written as 
\begin{equation} 
\left|\rmT\right|^2=  \left|\rmT_{\rmDVCS}\right|^2  
   +  \left|\rmT_{\rmBH}\right|^2
   +  \underbrace{\rmT_{\rmDVCS}\rmT_{\rmBH}^*
                 +\rmT_{\rmDVCS}^*\rmT_{\rmBH}
                }_{\rmI},
\label{eq:eq1}
\end{equation}
where `$\rmI$' denotes the interference term. 
The BH amplitude $\rmT_\rmBH$ is precisely calculable from measured elastic form factors of the nucleon and provides the dominant contribution in Eq.~\protect\ref{eq:eq1} in the kinematic conditions of the present measurement.
These amplitudes depend on $Q^2= -q^2$ with $q= k-k^\prime$ and $k$ ($k^\prime$) the four-momentum of the lepton in the initial (final) state, the variable $x_B= Q^2/(2 M\nu)$ with $\nu= p\cdot q/M$ and $M$ the nucleon mass, and $t$.
%, and the squared four-momentum transfer $t=(p-p^\prime)^2$ 
%to the nucleon with $p$ ($p^\prime$) the four-momentum of the nucleon in the intial (final) state. 
In addition, the amplitudes depend on $\phi$ and, in the case of a target polarisation component orthogonal to $\vec{q}$, on $\phi_S$, the azimuthal angles about the virtual-photon direction that are defined in Fig.~\ref{fig:coordinate}. The dependences on $\phi$ related to beam helicity 
and beam charge have been investigated experimentally~\cite{Air01,Stepanyan:2001sm,HallAxsec,Air07}, 
resulting in first constraints on GPDs.

At Leading Order (LO) in the fine-structure constant $\alpha_{em}$, the squared BH amplitude $\left|\rmT_{\rmBH}\right|^2$ is independent of the target polarisation with an unpolarised beam, and independent of the lepton charge. 
In contrast, the squared DVCS amplitude $\left|\rmT_{\rmDVCS}\right|^2$ and the interference term $\rmI$ can depend on the target polarisation even with an unpolarised beam, and the sign of the interference term depends also on the lepton charge. 
For an unpolarised lepton beam and a transversely polarised nucleon target, these dependences 
read~\cite{gous,Bel02a}
\begin{eqnarray} 
\left|\rmT_{BH}\right|^2 &=& \frac{K_\rmBH}{\pgate}
\Bigl(c_{0,\rmunp}^\rmBH+
\Lbrase c_{1,\rmunp}^\rmBH\cos\phi
+\lbrase c_{2,\rmunp}^\rmBH\cos(2\phi)\rbrase \Rbrase\Bigr),\label{eq:BH2} \\%[0.2cm]
\left|\rmT_{DVCS}\right|^2 &=& K_\rmDVCS\Bigl(c_{0,\rmunp}^\rmDVCS+c_{2,\rmunp}^\rmDVCS\cos(2\phi) 
+ \lbrase c_{1,\rmunp}^\rmDVCS\cos\phi \rbrase \Bigr.\nonumber \\
&&+S_\perp\Bigl[\bm{c_{0,\rmTP}^\rmDVCS\sin{}(\phi-\phi_S)}+c_{2,\rmTP}^\rmDVCS\sin(\phi-\phi_S)\cos(2\phi)\Bigr.\nonumber\\
&&\hspace{.5cm}+s_{2,\rmTP}^\rmDVCS\cos(\phi-\phi_S)\sin(2\phi) \label{eq:DVCS2} \\ 
&&\hspace{.4cm}\Bigl.\Bigl.+\lbrase c_{1,\rmTP}^\rmDVCS\sin(\phi-\phi_S)\cos\phi+s_{1,\rmTP}^\rmDVCS\cos(\phi-\phi_S)\sin\phi\rbrase\Bigr]\Bigr), \nonumber\\[0.2cm]
%\end{eqnarray}
%\begin{eqnarray}
\hspace{1.4cm}\rmI &=& \frac{-K_\rmI e_l}{\pgate}\Bigl(\bm{c_{1,\rmunp}^\rmI\cos{}\phi}+c_{3,\rmunp}^\rmI\cos(3\phi)\Bigr.\nonumber\\
&&\hspace{3cm}\left.+\lbrase \bm{c_{0,\rmunp}^\rmI}+c_{2,\rmunp}^\rmI\cos(2\phi)\rbrase\right.\nonumber\\
&&+S_\perp\Bigl[\bm{c_{1,\rmTP}^\rmI\sin{}(\phi-\phi_S)\cos{}\phi}+\bm{s_{1,\rmTP}^\rmI\cos(\phi-\phi_S)\sin{}\phi}\Bigr.\nonumber\\
&&\hspace{0.5cm}+c_{3,\rmTP}^\rmI\sin(\phi-\phi_S)\cos(3\phi)+s_{3,\rmTP}^\rmI\cos(\phi-\phi_S)\sin(3\phi)\nonumber\\
&&\hspace{0.5cm}+\lbrase\bm{c_{0,\rmTP}^\rmI\sin{}(\phi-\phi_S)}+c_{2,\rmTP}^\rmI\sin(\phi-\phi_S)\cos(2\phi)\right.\nonumber\\
&&\hspace{0.8cm}\Bigl.\Bigl.\left.+s_{2,\rmTP}^\rmI\cos(\phi-\phi_S)\sin(2\phi)\rbrase\Bigr]\Bigr).\label{eq:I}
\end{eqnarray} 
Here, $S_\perp$ denotes the magnitude of the transverse target polarisation, $e_l$ the beam charge in units of the elementary charge, $\pgate$ contains the $\phi$-dependent lepton propagators, and the braces enclose terms that are kinematically suppressed by $1/Q$. 
The subscripts `$\rmunp$' and `$\rmTP$' denote an unpolarised beam with unpolarised and transversely polarised targets, respectively. 
The dependences of the coefficients $c_n$ and $s_n$ on GPDs are elaborated in Ref.~\cite{Bel02a},
where the kinematic factors $K$ are defined. 
The factor $K_\rmDVCS$ in Eq.~\protect\ref{eq:DVCS2} suppresses the squared DVCS amplitude by two orders of magnitude
relative to the interference term in the kinematics of the present measurement. 
Note that the azimuthal angles defined here are different from those used in Ref.~\cite{Bel02a} 
($\phi=\pi-\phi_{[15]}
%{\cite{Bel02a}}
$ and $\phi-\phi_s=\pi+\varphi_{[15]}
%{\cite{Bel02a}}
$), leading to opposite 
signs for some of the coefficients given below.

The terms of particular interest in this paper appear in bold face in Eqs.~\ref{eq:DVCS2} and \ref{eq:I}.
The corresponding coefficients
%\footnote{Check the signs} 
can be approximated as 
\begin{eqnarray} 
c_{0,\rmTP}^\rmDVCS&\propto&-\frac{\sqrt{-t}}{M}\rmIm\left\{\bm{\CalH\CalE^*-\CalE\CalH^*
+\xi\CalEtil\,\CalHtil^*-\CalHtil\,\xi\CalEtil^*}\right\}, \label{eq:cT0DVCS}\\ 
c_{1,\rmunp}^\rmI&\propto&\frac{\sqrt{-t}}{Q}\rmRe\left\{\bm{F_1\CalH} 
+ \xi(F_1+F_2)\CalHtil - \frac{t}{4M^2}F_2\CalE \right\}, \label{eq:cU1I}\\ 
 c_{0,\rmunp}^\rmI&\propto&-\frac{\sqrt{-t}}{Q} c_{1,\rmunp}^\rmI, \label{eq:cU0I} \\
c_{1,\rmTP}^\rmI&\propto&-\frac{M}{Q}\rmIm\left\{\bm{\frac{t}{4M^2}\left[(2-x_B)F_1\CalE-4\frac{1-x_B}{2-x_B}F_2\CalH\right]} \right. \nonumber \\
&&\hspace*{12mm}  \left. + x_B\xi\left[F_1(\CalH+\CalE)-(F_1+F_2)(\CalHtil+\frac{t}{4M^2}\CalEtil)\right]\right\}, \label{eq:cT1I} \\
c_{0,\rmTP}^\rmI &\propto& -\frac{\sqrt{-t}}{Q} c_{1,\rmTP}^\rmI,  \label{eq:cT0I} \\
s_{1,\rmTP}^\rmI&\propto&-\frac{M}{Q} \rmIm\left\{\bm{\frac{t}{4M^2}\left[
   4\frac{1-x_B}{2-x_B}F_2\CalHtil-(F_1+\xi F_2)x_B\CalEtil}\right]\right. 
\label{eq:sT1I} \\
&+& \left.x_B\left[(F_1+F_2)\left(\xi\CalH+\frac{t}{4M^2}\CalE\right)-\xi F_1(\CalHtil+\frac{x_B}{2}\CalEtil)\right]\right\}, \nonumber
\end{eqnarray} 
with the skewness $\xi$ approximated by $\xi\approx x_B/(2-x_B)$. 
The Compton form factors $\CalH$, $\CalE$, $\CalHtil$ and $\CalEtil$ are convolutions of hard scattering amplitudes with the corresponding twist-two quark GPDs $H_q$, $E_q$, $\widetilde{H}_q$ and $\widetilde{E}_q$, while $F_1$ and $F_2$ are the nucleon Dirac and Pauli form factors~\cite{Bel02a}. 
In Eqs.~\protect\ref{eq:cT0DVCS}-\ref{eq:sT1I}, the use of  bold face differs from that in
Eqs.~\protect\ref{eq:DVCS2} and \ref{eq:I}.
Here, the terms not in bold face are suppressed relative to those in bold face in the same equation by either $x_B$ (or $\xi$) or $t/M^2$, which are of order 0.1 in the kinematic conditions of this measurement.  
The terms containing $x_B \CalEtil$ (or $\xi \CalEtil$) are not suppressed 
because the pion-pole contribution to 
$\widetilde{E}$ scales as $1/x_B$.

The coefficients $c_{0,\rmunp}^\rmI$ and $c_{1,\rmunp}^\rmI$ provide an 
experimental constraint on the real part of the Compton form factors, 
and can be used to test various models for GPDs as in Ref.~\cite{Air07}. 
Most importantly for the present work, the coefficients $c_{0,\rmTP}^\rmDVCS$, $c_{0,\rmTP}^\rmI$ and $c_{1,\rmTP}^\rmI$ provide rare access to the GPD $E$ with no kinematic suppression of its contribution relative to those of the other GPDs.
% As the total angular momentum $J_q$ carried by quarks of flavor $q$ in the nucleon is given by 
% the $x$-second moment of $H_q+E_q$ in the limit $t\rightarrow 0$~\cite{Ji97}, 
Measurements sensitive to these coefficients may provide via the Ji relation (Eq.~\ref{eq:JiRelation})
an opportunity to constrain parameterisations of the GPD $E_q$ in terms of $J_q$~\cite{Elli05}.
%\footnote{Ellinghaus should insert a sentence on Eq.~10.}
The coefficient $c_{0,\rmunp}^\rmI$ ($c_{0,\rmTP}^\rmI$) has approximately the same dependence on GPDs as $c_{1,\rmunp}^\rmI$ ($c_{1,\rmTP}^\rmI$).
The apparent overall suppression of $c_{0,\rmunp}^\rmI$ and $c_{0,\rmTP}^\rmI$ by $\sqrt{-t}/Q$ with respect to $c_{1,\rmunp}^\rmI$ and $c_{1,\rmTP}^\rmI$ is compensated by an enhancement from $y$-dependent factors that are not shown, where $y=p\cdot q/p\cdot k$. 
These factors range from two to four in the kinematic conditions of the present measurement.
The previously mentioned strong kinematic suppression of the squared DVCS amplitude relative to the interference term is partially compensated by the unshown kinematic factors 
that apply to Eqs.~\ref{eq:cT0DVCS}, \ref{eq:cT1I} and \ref{eq:cT0I}.
The net suppression is only about one order of magnitude  in the HERMES kinematic conditions, and some sensitivity to the GPD $E$ may therefore be provided by $c_{0,\rmTP}^\rmDVCS$.
The coefficient $s_{1,\rmTP}^\rmI$ provides experimental sensitivity to the 
GPD $\widetilde{E}$, and also to  $\widetilde{H}$,
which was already probed experimentally through measurements of longitudinal target-spin asymmetries \cite{kopyt,Jlabhallb2006}.
The coefficients $c_{2,\rmTP}^\rmDVCS$, $s_{2,\rmTP}^\rmDVCS$, $c_{3,\rmunp}^\rmI$, $c_{3,\rmTP}^\rmI$ and $s_{3,\rmTP}^\rmI$ receive twist-two contributions involving the unknown gluon helicity-flip GPDs~\cite{Die01}.
These GPDs do not mix with quark GPDs via $Q^2$ evolution 
and thus probe the intrinsic gluonic 
properties of the nucleon \cite{Bel00}.
As the contribution of gluon helicity-flip is suppressed by the strong coupling constant $\alpha_S$, this contribution competes with that from twist-four quark GPDs, which is suppressed by a factor $M^2/Q^2$ but not by $\alpha_s$~\cite{Kiv01}.
Aside from $c_{0,\rmunp}^\rmDVCS$ and $c_{2,\rmunp}^\rmDVCS$,
all other coefficients appearing in
Eqs.~\protect\ref{eq:DVCS2} and \ref{eq:I} are related to twist-three quark GPDs.
%% Dietmar (Sep 2): We discussed footnote 1 two weeks ago. If Frank has no objections, I think we can delete it.
%% \footnote{Dietmar and Frank E. will check this}

\section{The experiment}

Hard exclusive production of real photons in the reaction $ep^\uparrow \rightarrow e^\prime \gamma p^\prime$ is studied.
Data with a transversely polarised hydrogen target~\cite{Air05} were accumulated using the HERMES spectrometer~\cite {Ack98} and the longitudinally polarised 27.6 GeV electron and positron beams of the HERA accelerator at DESY. 
This final data set with the transversely polarised target 
was collected over the years 2002-2005. 
The integrated luminosities for the electron and positron samples are approximately 100 pb$^{-1}$ and 70 pb$^{-1}$, respectively.

Events are selected if there were detected exactly one photon and one charged track identified as the scattered lepton. 
%In this sample  leptons are contaminated by misidentified hadrons to less than 1\%
The hadron contamination in the lepton sample is kept below 1\% 
by combining the information from a transition-radiation detector, a preshower scintillator detector, and an electromagnetic calorimeter. 
The kinematic requirements imposed are 1~GeV$^2 < Q^2 <$ 10~GeV$^2$, 0.03 $< x_B <$ 0.35, 
and $\nu <$ 22 GeV.
The real photon is identified through the appearance of a `neutral signal cluster', which is defined as an energy deposition larger than 5 GeV in the calorimeter with a signal larger than 1 MeV in the preshower detector, and the absence of a corresponding charged track in the back region of the spectrometer. 
The angular separation $\theta_{\gamma^* \gamma}$ between the virtual and real photons is required to be larger than 5 mrad. This value is determined mainly by the lepton momentum resolution. An upper bound of 45 mrad
is imposed  on this angle in order to improve the signal-to-background ratio~\cite{Ell04}.

\begin{figure}
\begin{center}
\includegraphics[width=0.8\columnwidth]{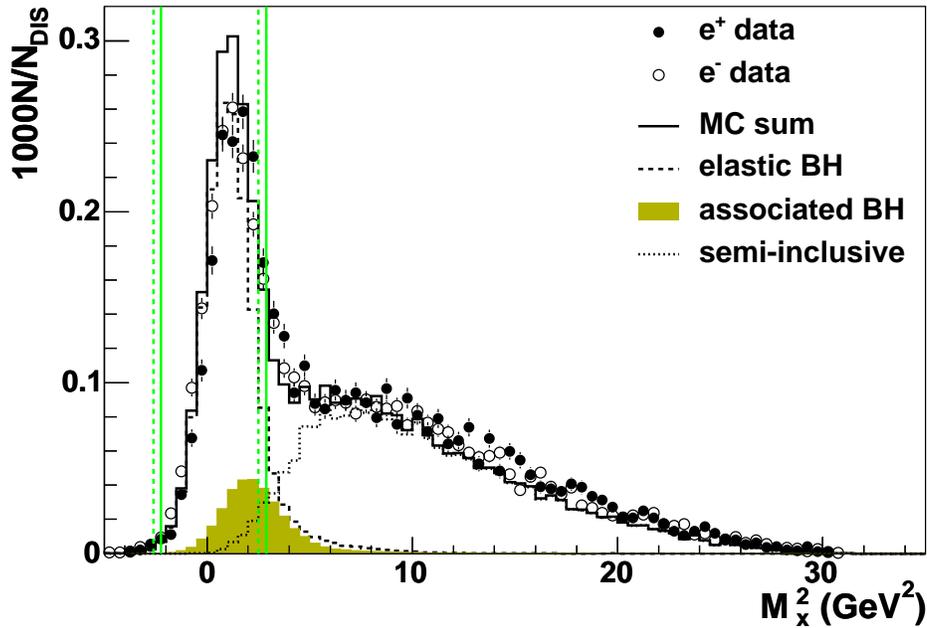}
\end{center}
\caption{Distributions in squared missing-mass from data with positron (filled points) and electron (empty circles) beams and from Monte Carlo simulations (solid line). 
The latter include elastic BH (dashed line) and associated BH (filled area) processes as well as semi-inclusive background (dotted line). 
The simulations and data are both absolutely normalized.
The vertical solid (dashed) lines enclose the selected exclusive region for the positron (electron) data. See text for details.
\label{missmass2}}
\end{figure}

The recoiling proton was not detected.
Instead, an `exclusive' sample of events is selected by requiring the squared missing mass $M_X^2= ( q + p - q^\prime)^2$ of the reaction $e p \rightarrow e^\prime\gamma X$ to be close to the squared proton mass, where $q^\prime$ is the four-momentum of the real photon. 
The selection criterion was chosen by means of a Monte Carlo (MC) simulation
%~\cite{Ye06}
of the distribution in $M_X^2$.
The simulation is shown in comparison with experimental data in Fig.~\ref {missmass2}.
%, where the positron and electron data and various contributions in the simulation are shown separately.
In the MC simulation \cite{Bernhard}, the Mo-Tsai formalism~\cite{Mo69} is used for the elastic BH process that leaves the target nucleon intact, as well as the BH process where the nucleon is excited to a resonant state (a category known as associated production). 
For the latter, a parameterisation of the total $\gamma^*p$ cross section for the resonance region is used~\cite {Bra76}. 
The individual cross sections for 
single-meson decay channels, e.g., $\Delta^+\rightarrow p \pi^0$, 
are treated according to the {\sc Maid}2000 model~\cite{maid2000}. The remaining contribution is assigned to 
multi-meson decay channels, e.g., $\Delta^+ \rightarrow p\pi^0\pi^0$, whose relative contributions are determined according to isospin relations.
The simulation also takes into account the semi-inclusive production of neutral mesons (mostly $\pi^0$) where either only one decay photon is detected or the decay photons cannot be resolved. 
For this, the MC generator {\sc Lepto}~\cite {Ing97} is used in conjunction with a 
set of  {\sc Jetset}~\cite{Sjo94} fragmentation parameters that had previously been adjusted
to reproduce multiplicity distributions observed by HERMES~\cite{Hil05}. 
Not shown in Fig.~\ref {missmass2} is the contribution from exclusive $\pi^0$ production, which is found to be less than 0.5\% in the exclusive region using the model in Ref.~\cite {Van99}. 
The MC yield exceeds the data by about 20\% in the exclusive region. This may be due to 
radiative effects not included in the simulation, which would move events from the peak to the continuum and improve the agreement~\cite {Van00}.
On the other hand, if the DVCS process were included in the simulation,
its contribution would increase the elastic peak.
This contribution is highly model-dependent and can vary between 10 and 25\%~\cite{Ye06}. 
The exclusive region for the positron data 
is chosen to be $-(1.5$~GeV)$^2$ $< M_X^2 < (1.7$~GeV)$^2$, 
where the value of $-(1.5$~GeV)$^2$ is displaced from the 
squared proton mass by three times the resolution in $M_X^2$, 
and the value of (1.7~GeV)$^2$ is the point where the 
contributions from the signal and background are equal. 
As the $M_X^2$ spectrum of the electron data is found to be shifted by approximately 0.18~GeV$^2$ towards smaller values relative to that of the positron data, 
% due to imperfections in the time dependence of the calorimeter calibration, 
the exclusive region for the electron data is shifted accordingly.
One quarter of the effect of this shift on the results presented below is assigned as a systematic uncertainty contribution.

As the recoiling proton remains undetected, $t$ is inferred from the measurement of the other final-state particles. 
For elastic events (leaving the proton intact), the kinematic relationship between the energy and direction of the real photon permits $t$ to be calculated without using the measured energy of the real photon, which is the quantity subject to larger uncertainty. 
Thus the value of $t$ in the exclusive region is calculated as 
\begin{equation}
t = \frac{-Q^2 - 2 \, \nu \, (\nu - \sqrt{\nu^2 + Q^2} \, \cos\theta_{\gamma^* \gamma })}
{1 + \frac{1}{M} \, (\nu - \sqrt{\nu^2 + Q^2} \, cos\theta_{\gamma^* \gamma })}\label{tc},
\end{equation}
which is exact for elastic events.
Using this method, the average resolution (RMS) in $t$ is improved from 0.11 to 0.03 GeV$^2$. Exclusive events are selected with $-t<$ 0.7 GeV$^2$ in order to reduce background. 

\section{Azimuthal asymmetries}

Experimental observables that provide sensitivity to the coefficients appearing in Eqs.~\ref{eq:DVCS2} and \ref{eq:I} are the Beam-Charge Asymmetry (BCA) 
\begin{equation}
\CalAC(\phi)\equiv\frac{d\sigma^+(\phi)-d\sigma^-(\phi)}{d\sigma^+(\phi)+d\sigma^-(\phi)}, \label{eq:BCA}
\end{equation}
and the Transverse Target-Spin Asymmetries (TTSAs) 
\begin{eqnarray}
\hspace{-2.cm}\CalAUTDVCS(\phi,\phi_S)&\equiv& \nonumber \\
&&\hspace{-1cm}\frac{1}{S_\perp}\cdot\frac{d\sigma^+(\phi,\phi_S)-d\sigma^+(\phi,\phi_S+\pi)+d\sigma^-(\phi,\phi_S)-d\sigma^-(\phi,\phi_S+\pi)}{d\sigma^+(\phi,\phi_S)+d\sigma^+(\phi,\phi_S+\pi)+d\sigma^-(\phi,\phi_S)+d\sigma^-(\phi,\phi_S+\pi)},\label{eq:TTSA1} \\
%\end{multline}
%\begin{multline}
\hspace{-2.cm}\CalAUTI(\phi,\phi_S)&\equiv& \nonumber \\
&&\hspace{-1cm}\frac{1}{S_\perp}\cdot\frac{d\sigma^+(\phi,\phi_S)-d\sigma^+(\phi,\phi_S+\pi)-d\sigma^-(\phi,\phi_S)+d\sigma^-(\phi,\phi_S+\pi)}{d\sigma^+(\phi,\phi_S)+d\sigma^+(\phi,\phi_S+\pi)+d\sigma^-(\phi,\phi_S)+d\sigma^-(\phi,\phi_S+\pi)}.\label{eq:TTSA2}
\end{eqnarray}
Here the subscripts on the $\mathcal{A}$'s represent dependence on 
beam Charge (C) or Transverse (T) target polarisation, 
with an Unpolarised (U) beam, and 
the superscripts $\pm$ stand for the lepton beam charge.
These asymmetries are related to the coefficients 
in  Eqs.~\ref{eq:BH2}--\ref{eq:I} by:
\begin{eqnarray}
\hspace{-2.4cm}\CalAC(\phi)&=& 
% &\rule{0mm}{1mm}&
\frac{-\frac{K_\rmI}{{\mathcal P}_1(\phi){\mathcal P}_2(\phi)} \sum_{n=0}^3 c_{n,\rmunp}^I \cos(n\phi)}{\frac{K_\rmBH}{{\mathcal P}_1(\phi){\mathcal P}_2(\phi)}\sum_{n=0}^2 c_{n,\rmunp}^\rmBH \cos(n\phi)+ K_\rmDVCS \sum_{n=0}^2 c_{n,\rmunp}^\rmDVCS \cos(n\phi)} \\
&\simeq& \frac{-K_\rmI\Bigl[
%c_{0,\rmunp}^I + 
c_{1,\rmunp}^I \cos(\phi)\Bigr]}{K_\rmBH c_{0,\rmunp}^\rmBH}, 
\end{eqnarray}
\begin{eqnarray}
\hspace{-2.4cm}\CalAUTDVCS(\phi,\phi_S)&=& \nonumber \\
&&\hspace{-2.4cm}\frac{K_\rmDVCS\Bigl[\sum_{n=0}^2 c_{n,\rmTP}^\rmDVCS\sin(\phi-\phi_S)\cos(n\phi) +\sum_{n=1}^2 s_{n,\rmTP}^\rmDVCS\cos(\phi-\phi_S)\sin(n\phi) \Bigr]}{\frac{K_\rmBH}{{\mathcal P}_1(\phi){\mathcal P}_2(\phi)}\sum_{n=0}^2 c_{n,\rmunp}^\rmBH \cos(n\phi)+ K_\rmDVCS \sum_{n=0}^2 c_{n,\rmunp}^\rmDVCS \cos(n\phi)},\label{eq:AUTDVCS} \\
&\simeq& \frac{K_\rmDVCS c_{0,\rmTP}^\rmDVCS\sin(\phi-\phi_S)}{\frac{K_\rmBH}{{\mathcal P}_1(\phi){\mathcal P}_2(\phi)} c_{0,\rmunp}^\rmBH} \\
%\end{multline}
%\begin{multline}
\hspace{-2.4cm}\CalAUTI(\phi,\phi_S)&=& \nonumber \\
&&\hspace{-2.4cm}\frac{- \frac{K_\rmI e_l}{{\mathcal P}_1(\phi){\mathcal P}_2(\phi)} \Bigl[ \sum_{n=0}^3 c_{n,\rmTP}^\rmI\sin(\phi-\phi_S)\cos(n\phi) + \sum_{n=1}^3 s_{n,\rmTP}^\rmI\cos(\phi-\phi_S)\sin(n\phi) \Bigr]}{\frac{K_\rmBH}{{\mathcal P}_1(\phi){\mathcal P}_2(\phi)}\sum_{n=0}^2 c_{n,\rmunp}^\rmBH \cos(n\phi)+ K_\rmDVCS \sum_{n=0}^2 c_{n,\rmunp}^\rmDVCS \cos(n\phi)}\label{eq:AUTI}\\
&\simeq& \frac{- K_\rmI e_l \Bigl[c_{1,\rmTP}^\rmI\sin(\phi-\phi_S)\cos\phi + s_{1,\rmTP}^\rmI\cos(\phi-\phi_S)\sin(\phi) \Bigr]}{K_\rmBH c_{0,\rmunp}^\rmBH}.
\end{eqnarray}
HERMES reported the first measurement of the BCA~\cite{Air07}, providing access to $c_{1,\rmunp}^\rmI$, the coefficient of the $\cos\phi$ modulation of the interference term for an unpolarised target. 
The present paper reports more precise BCA results using a considerably larger new data set from the first DVCS measurement done with transverse target polarisation.
Most importantly, the extracted TTSAs provide access to $c_{0,\rmTP}^\rmDVCS$, $c_{0,\rmTP}^\rmI$ and $c_{1,\rmTP}^\rmI$, which are sensitive to the total angular momentum of quarks in the nucleon, as noted above. 
%These results are extracted in a combined fit that separates information related to the squared DVCS amplitude and interference term.

\subsection{Extraction of azimuthal asymmetry amplitudes} 

The distribution in the expectation value of the yield is given by:
\begin{eqnarray}
\hspace{-1.5cm}\langle\intN\rangle(S_\perp,e_l,\phi,\phi_S)&=&\Lumi(S_\perp,e_l)\,\epsilon(e_l,\phi,\phi_S)\,\CUU(\phi) \nonumber \times \\
&\rule{0mm}{1mm}&\Bigl[1+S_\perp\CalAUTDVCS(\phi,\phi_S)+e_l\CalAC(\phi)+e_lS_\perp\CalAUTI(\phi,\phi_S)\Bigr].\label{eq:A1}
\end{eqnarray}
Here $\Lumi$ is the integrated luminosity, $\epsilon$ the detection efficiency, 
and $\CUU$ the cross section for an unpolarised target averaged over both beam charges.
The BCA $\CalAC(\phi)$ and the TTSAs $\CalAUTDVCS(\phi,\phi_S)$ and $\CalAUTI(\phi,\phi_S)$ 
in Eq.~\ref{eq:A1} are expanded in terms of the same harmonics in $\phi$ and $\phi_S$ as those appearing in Eqs.~\ref{eq:DVCS2} and \ref{eq:I} (as well as the harmonics $\cos(\phi-\phi_S)\sin(3\phi)$ and $\sin(\phi-\phi_S)\cos(3\phi)$ in Eq.~\ref{eq:asymmetry2}, included as a systematic check):
\begin{eqnarray}
\CalAC(\phi;\ttC)&=&\sum_{n=0}^3 \AC^{\cos(n\phi)}\cos(n\phi),\label{eq:asymmetry1} 
\end{eqnarray}
\begin{eqnarray}
\CalAUTDVCS(\phi,\phi_S;\ttDVCS)&=&\sum_{n=0}^3 \AUTDVCS^{\sin(\phi-\phi_S)\cos(n\phi)}\sin(\phi-\phi_S)\cos(n\phi)\nonumber\\
&&+\sum_{n=1}^3 \AUTDVCS^{\cos(\phi-\phi_S)\sin(n\phi)}\cos(\phi-\phi_S)\sin(n\phi),\label{eq:asymmetry2} 
\end{eqnarray}
\begin{eqnarray}
\CalAUTI(\phi,\phi_S;\ttI)&=&\sum_{n=0}^3 \AUTI^{\sin(\phi-\phi_S)\cos(n\phi)}\sin(\phi-\phi_S)\cos(n\phi)\nonumber\\
&&+\sum_{n=1}^3 \AUTI^{\cos(\phi-\phi_S)\sin(n\phi)}\cos(\phi-\phi_S)\sin(n\phi).\label{eq:asymmetry3} 
\end{eqnarray}
Here $\ttC$, $\ttDVCS$ and $\ttI$ represent the sets of Fourier coefficients or azimuthal asymmetry amplitudes, 
hereafter called `asymmetry amplitudes',
% $\AUTDVCS$, $\AC$ and $\AUTI$ 
appearing in the right-hand sides of Eqs.~\ref{eq:asymmetry1}--\ref{eq:asymmetry3}
describing respectively the dependences of the squared DVCS amplitude and interference term on beam Charge (C), Transverse (T) target polarisation
 or both, with an Unpolarised (U) beam.
These 18 asymmetry amplitudes embody the essential sensitivities to GPD models of the coefficients of the corresponding functions of $\phi$ appearing in Eqs.~\ref{eq:DVCS2} and \ref{eq:I}, to the degree that one can neglect the effects of the coefficients $c_{1,\rmunp}^\rmBH$ and $c_{2,\rmunp}^\rmBH$ and the squared unpolarised DVCS amplitude in Eqs.~\ref{eq:AUTDVCS}--\ref{eq:AUTI} and the effect of the $\phi$-dependence of the BH propagators.
In any case, the extracted asymmetry amplitudes are well defined and can be computed in various GPD models for direct comparison with the data.
For each kinematic bin in $-t$, $x_B$ or $Q^2$, they are simultaneously extracted from the observed exclusive event sample using the method of Maximum Likelihood.
The distribution of events is parameterised by the function $\intNpar$, which is defined as
\begin{multline}
\intNpar(S_\perp,e_l,\phi,\phi_S;\ttDVCS,\ttC,\ttI)=\Lumi(S_\perp,e_l)\,\epsilon(e_l,\phi,\phi_S)\,\CUU(\phi) \times\\
\Bigl[1+S_\perp\CalAUTDVCS(\phi,\phi_S;\ttDVCS)+e_l\CalAC(\phi;\ttC)+e_lS_\perp\CalAUTI(\phi,\phi_S;\ttI)\Bigr].\label{eq:distribution}
\end{multline}
While the net beam polarisations of both positron and electron data samples used in the current measurement are not completely negligible ($0.03\pm0.02$ and $-0.03\pm0.02$, respectively), 
algebraic investigations (confirmed by MC studies) show that this does not affect the asymmetry amplitudes presented here.
Not included in Eqs.~\ref{eq:asymmetry1}-\ref{eq:asymmetry3} are negligible terms involving the small component of the target polarisation that is parallel to $\vec{q}$~\cite{Die05}.

Within the scheme known as Extended Maximum Likelihood~\cite{Bar90}, the likelihood function $L$ to be minimized is taken as
\begin{eqnarray}
-\ln{L(\ttDVCS,\ttC,\ttI)}&=&\intNpartil(\ttDVCS,\ttC,\ttI)
\nonumber\\
&&\hspace{-3cm}-\sum_{i=1}^{\Nobs}\ln \Bigl[1+S_\perp^i\CalAUTDVCS(\phi^i,\phi_S^i;\ttDVCS)+e_l^i\CalAC(\phi^i;\ttC)\nonumber\\
&&\hspace{-1cm}+e_l^iS_\perp^i\CalAUTI(\phi^i,\phi_S^i;\ttI)\Bigr],\label{eq:Likelihood}
\end{eqnarray}
where $\Nobs$ is the observed number of events, and the parameterised total number $\intNpartil$ of events is 
\begin{eqnarray}
\intNpartil(\ttDVCS,\ttC,\ttI)&=& \nonumber \\
&&\hspace{-2cm}\int \de S_\perp\,d\phi\,d\phi_S\sum_{e_l=\pm1}\intNpar(S_\perp,e_l,\phi,\phi_S;\ttDVCS,\ttC,\ttI).\label{eq:expect}
\end{eqnarray}
%Here $K_\rmBH$, $K_\rmDVCS$ and $K_\rmI$ stand for kinematic factors that are not shown in Eqs.~\ref{eq:BH2}-\ref{eq:I}.
The cross section $\CUU$ and the detection efficiency $\epsilon$ do not depend on $\ttDVCS$, $\ttC$ or $\ttI$ and thus cannot affect the location of the likelihood maximum.
%\footnote{Eduard: take Gunar's GMC, run MC with t-magnet effect}
Hence they have been omitted in the logarithms in Eq.~\ref{eq:Likelihood}. 
It is also not necessary to consider them explicitly
in evaluating $\intNpartil(\ttDVCS,\ttC,\ttI)$ in Eq.~\ref{eq:expect},
because the needed information about them is encoded in the
total yields obtained by combining events for both target 
polarisations and beam charges.  
Luminosity imbalances between beam charges or target polarisations
are taken into account by assigning 
weights $w_i$ to the events, which are adjusted to 
provide effectively vanishing net target polarisation and 
net beam charge for this combined data set.
The weights are normalized to also retain the same integrated luminosity
$\Lumi_{tot}$ as the observed `unweighted' data sample.
The resulting event distribution corresponds to the
product $\Lumi_{tot}\,\epsilon(\phi,\phi_S)\,\CUU(\phi)$.
In this manner, an event distribution
corresponding to Eq.~\ref{eq:distribution} is
constructed to estimate the parameterised total number of events in Eq.~\ref{eq:expect}:
%\begin{eqnarray}
% \hspace{-2cm}\intNpartil(\ttDVCS,\ttC,\ttI)&\approx&\sum_{i=1}^{\Nobs} w_i\,\Lumi(S_\perp,e_l)/\Lumi_{tot} \nonumber \\
% &&\hspace{-3.5cm}\times\Bigl[1+S_\perp^i\CalAUTDVCS(\phi^i,\phi_S^i;\ttDVCS)+e_l^i\CalAC(\phi^i;\ttC)+e_l^iS_\perp^i\CalAUTI(\phi^i,\phi_S^i;\ttI)\Bigr].
% \end{eqnarray}
\begin{multline}
\intNpartil(\ttDVCS,\ttC,\ttI)\approx\sum_{i=1}^{\Nobs} w_i\,\Lumi(S_\perp,e_l)/\Lumi_{tot} \nonumber \times\\
\Bigl[1+S_\perp^i\CalAUTDVCS(\phi^i,\phi_S^i;\ttDVCS)+e_l^i\CalAC(\phi^i;\ttC)+e_l^iS_\perp^i\CalAUTI(\phi^i,\phi_S^i;\ttI)\Bigr].
\end{multline}

\subsection{Background corrections and systematic uncertainties}

\begin{figure}[h!]
\includegraphics[width=1\columnwidth]{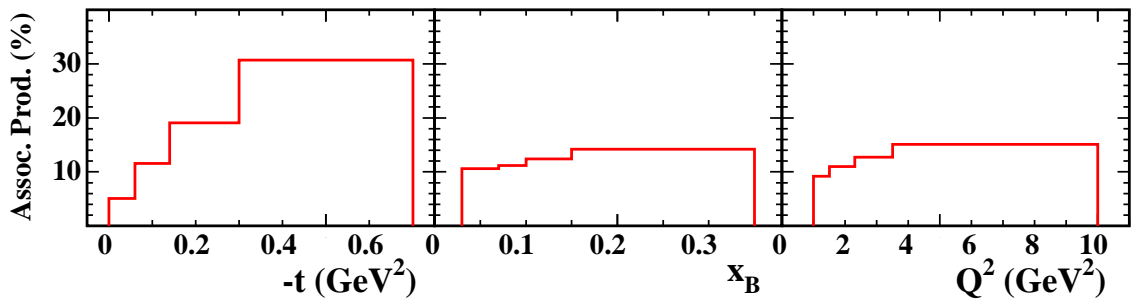}
\caption{Kinematic dependences of the simulated fractional contributions from  associated production. See text for details.}
\label{fig:frac}
\end{figure}

The results from the minimization of Eq.~\ref{eq:Likelihood} in each kinematic bin are corrected for background from semi-inclusive and exclusive production of neutral mesons, mainly pions, in order to estimate the true asymmetry amplitude:
\begin{equation}
A_t=\frac{A_r-s\cdot A_s-e\cdot A_e}{1-s-e},
\label{eq:corr}
\end{equation}
where $A_r$ stands for the extracted raw asymmetry amplitude, and
 $s$ and $A_s$ ($e$ and $A_e$) the fractional contribution and corresponding asymmetry amplitude of the semi-inclusive (exclusive) background. 
The combination of these background contributions $s+e$ ranges from $2\pm1\%$ to $11\pm5\%$ in the kinematic space~\cite{Ye06}. 
As these background contributions are only very weakly beam-charge dependent,
their asymmetries with respect to the beam charge or to the product of the beam charge and the transverse target polarisation are neglected.
The asymmetry of the semi-inclusive $\pi^0$ background with respect to only the transverse target polarisation is extracted from experimental data by requiring two `neutral signal clusters' in the calorimeter with their invariant mass between 0.10 and 0.17 GeV. 
The restriction on the energy deposition in the calorimeter of the less energetic neutral signal cluster is relaxed to 1 GeV to improve the statistical precision.
The fractional energy of the reconstructed neutral pions is required to be large, $z=E_\pi/\nu>0.8$, as only these contribute to the exclusive region according to MC simulations \cite{Ye06}.
These simulations showed that the extracted $\pi^0$ asymmetry does not depend on whether only one or both photons are in the acceptance. 
It is convenient to use the direction of the reconstructed pion  in place of that of the photon to calculate the azimuthal angles $\phi$ and $\phi_S$.
For the exclusive $\pi^0$ background, the asymmetry amplitudes with respect to only  target polarisation are not extracted 
due to the limited statistical precision but rather assumed to be $0\pm1$.
After applying  Eq.~\ref{eq:corr}, the resulting asymmetry amplitude $A_t$ is expected to originate from only  elastic and associated production. 
On average $12$\% of the BH cross section arises from the latter~\cite{Ye06}, according to the simulation 
described above. 
The kinematic dependences of this  contribution are shown in  Fig.~\ref{fig:frac}.
No correction is made or uncertainty assigned for associated production, 
as it is considered to be part of the signal.

The dominant contributions to the total systematic uncertainty are those from the detector acceptance and finite bin width, 
and the determination of the target polarisation~\cite{Ye06}.
The combined contribution to the systematic uncertainty 
from the detector acceptance, finite bin width, 
and the alignment of the detector elements with respect to the beam, including possible effects from the beam and track curvature in the transverse magnetic field of the target, is determined from MC simulations based on five %seven 
GPD models described in  Ref.~\cite{Kor02a}. % and \cite{Vad06}.
In each kinematic bin, it is defined as 
the RMS difference between the asymmetry amplitude extracted from the 
MC data in that bin by minimizing Eq.~\ref{eq:Likelihood}
and  the corresponding model predictions calculated analytically 
at the mean kinematic values of that bin given in Table~\ref{table:ttsa}.
The other sources are associated with the background correction, 
calorimeter calibration and the relative shift of the $M_X^2$ spectra between the positron and electron data.
These contributions, given in Table~\ref{table:syst}, are added in quadrature to form the total systematic uncertainty per kinematic bin, appearing in Table~\ref{table:ttsa}.
Not included is any contribution due to additional QED vertices, as the most significant of these has been estimated to be negligible~\cite{Afa06}.

\vspace*{-0.5cm}
\begin{figure}
\includegraphics[width=\columnwidth]{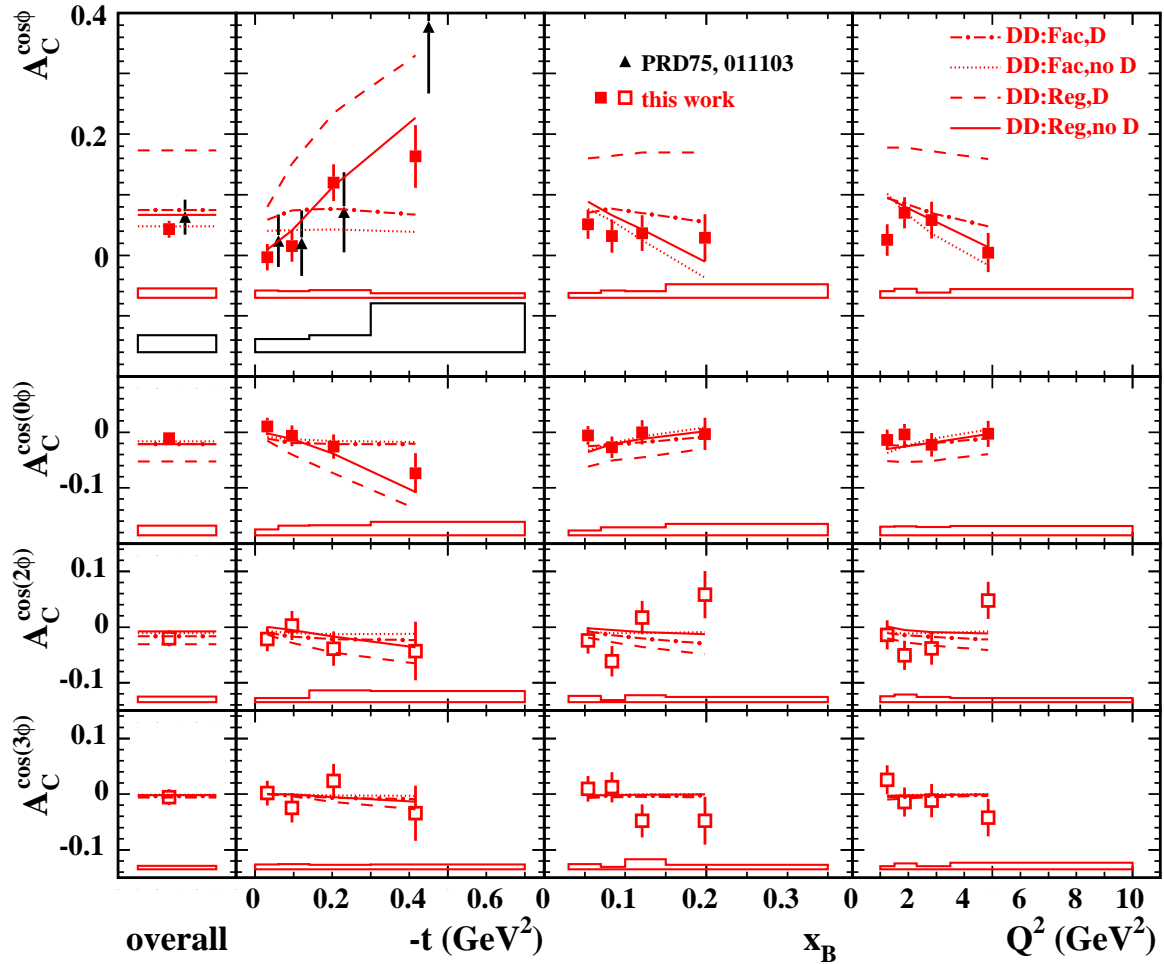}
\caption{Asymmetry amplitudes describing the dependence of the interference term 
on the beam charge ($\AC$), for the exclusive sample. 
The squares represent the results from the present work, while data represented 
by  triangles (shifted right for visibility) 
were reported in Ref.~\cite{Air07}. 
The filled symbols indicate those results of greatest interest (see text).
The error bars (bands) represent the statistical (systematic) uncertainties.
% The dominant systematic effect represented by the error band in the second row of panels ($\AC^{\cos(0\phi)}$) is independent of $-t$, $x_B$ or $Q^2$. 
The curves are  predictions of variants of a double-distribution GPD model~\cite{Van99,Goe01}, 
%with $b_v=\infty$ and $b_s=1$
with profile parameters given in Table~\ref{table:bprof}. 
See text for details.\label{fig:bca}}
\end{figure}

\begin{figure}
%\vspace*{1mm}
\includegraphics[width=1\columnwidth]{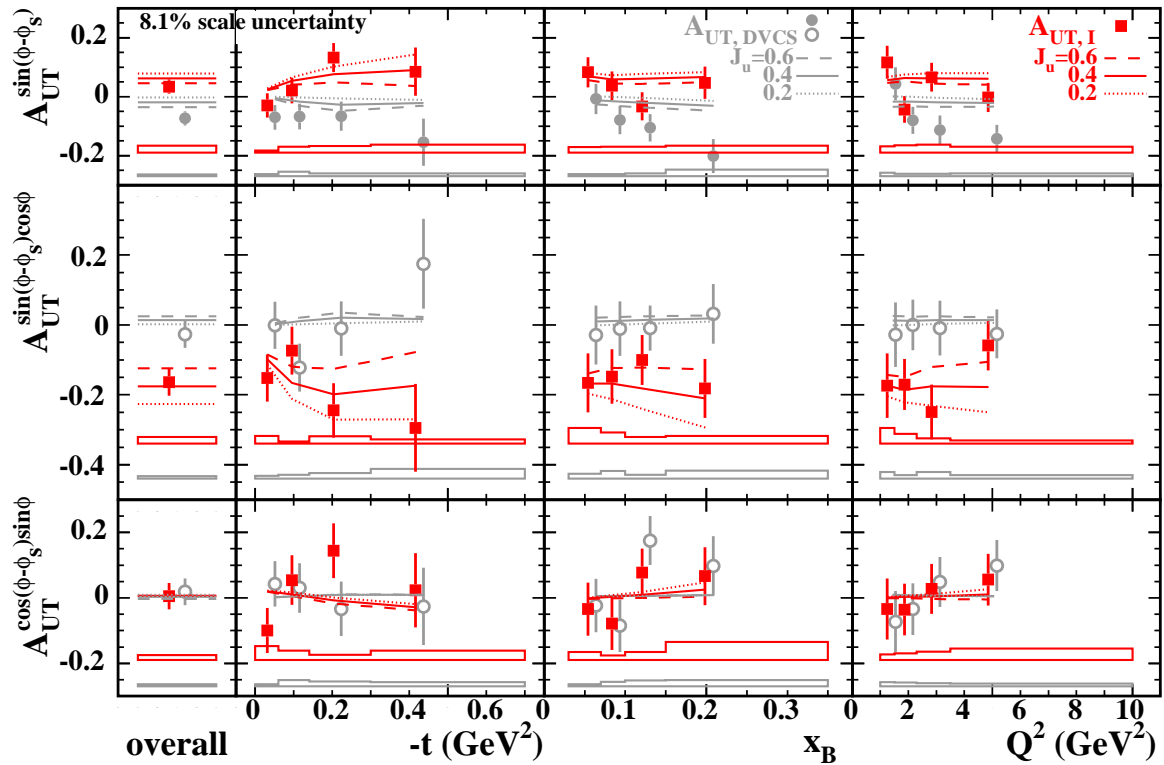}
\caption{Asymmetry amplitudes describing the dependence of the squared DVCS amplitude (circles, $\AUTDVCS$) and the interference term (squares, $\AUTI$) on the transverse target polarisation,
for the exclusive sample. 
The filled symbols indicate those results of greatest interest (see text).
The circles (squares) are shifted right (left) for visibility. 
The error bars represent the statistical uncertainties, while the top (bottom) bands denote the systematic uncertainties for $\AUTI$ ($\AUTDVCS$), excluding the 8.1~\% scale uncertainty from the target polarisation measurement.
The curves are predictions  of the
GPD model variant  (Reg, no D) shown in Fig.~\ref{fig:bca} as a continuous curve, with three different values for the $u$-quark total angular momentum $J_u$ and fixed $d$-quark total angular momentum $J_d=0$~\cite{Elli05}. 
See text for details.\label{fig:ttsa}}
\end{figure}

\begin{figure}
\vspace*{-1cm}
\includegraphics[width=1\columnwidth]{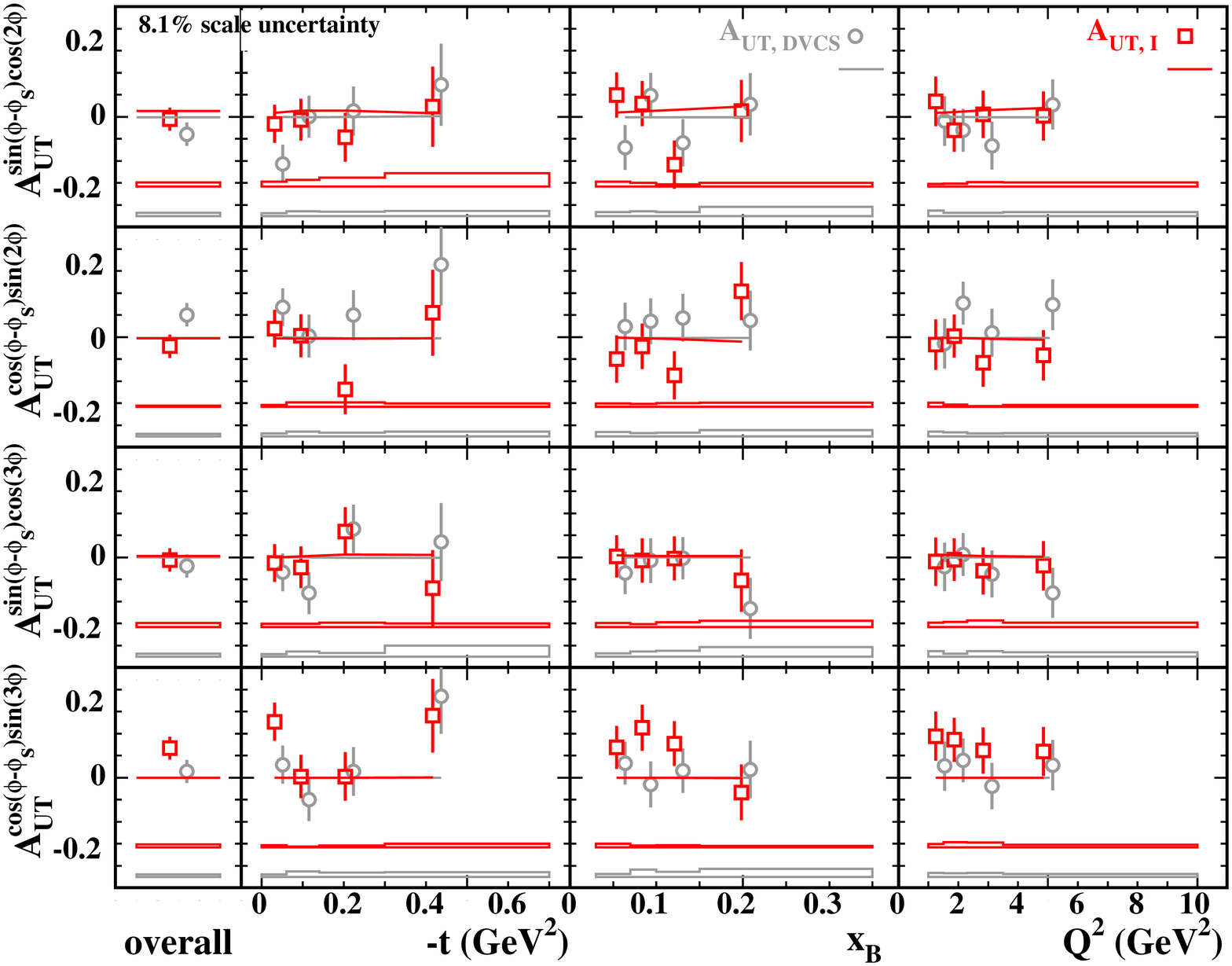}
\caption{Asymmetry amplitudes that are expected to be suppressed, presented as in Fig.~\ref{fig:ttsa}, except
that the curves are calculated only for $J_u=0.4$.
\label{fig:ttsa2}
}
\end{figure}

\section{Results}

Figures~\ref{fig:bca}--\ref{fig:ttsa2} show as a function of $-t$, $x_B$ or $Q^2$, in four bins, the results from the combined fit. 
The `overall' results in the left-most columns correspond to the entire experimental acceptance.
Fig.~\ref{fig:bca} shows the amplitudes related to only beam charge, while Figs.~\ref{fig:ttsa} and \ref{fig:ttsa2} show the amplitude $\AUTDVCS$, which relates to transverse target polarisation only, 
and $\AUTI$, which relates to both.
For simplicity of presentation, the amplitudes $\AUTDVCS$ and $\AUTI$ for the same azimuthal dependence are shown together in each panel, even though they typically do not relate to the same GPDs. 
The filled symbols represent the asymmetry amplitudes of interest here (see 
Table~\ref{table:ttsa}), related to the  coefficients given in Eqs.~\ref{eq:cT0DVCS}--\ref{eq:sT1I},
 of the corresponding harmonics of $\phi$ appearing in
Eqs.~\ref{eq:DVCS2} and \ref{eq:I}.

Of particular interest is the asymmetry amplitude $\AC^{\cos\phi}$ in the upper row of Fig.~\ref{fig:bca}.  Equation~\ref{eq:cU1I} shows that this amplitude is sensitive to the GPD $H$ in the HERMES kinematic conditions. 
Also shown in this figure is the previously published result, which has been shown to constrain GPD models~\cite{Air07}.
The greatly improved precision of the present measurement confirms that this amplitude increases with increasing $-t$.
As mentioned above 
regarding the corresponding coefficients $c_{0,\rmunp}^\rmI$ and $c_{1,\rmunp}^\rmI$,
the amplitude $\AC^{\cos0\phi}$ is expected to relate to the same combination of GPDs as does $\AC^{\cos\phi}$. 
The results shown in Fig.~\ref{fig:bca} suggest that the magnitude of this amplitude also increases with $-t$, while its opposite sign 
is expected from Eq.~\ref{eq:cU0I}.
%arises from the interplay between the coefficient $c_{1,\rmunp}^\rmI$ and the squared 
%BH amplitude $\left|\rmT_{\rmBH}\right|^2$.
%\footnote{Dietmar: Check if it is the case.  What does this ``interplay'' between a coefficient and an amplitude mean?}

Of special interest in this work are the amplitudes 
$\AUTI^{\sin(\phi-\phi_S)\cos(n\phi)}, n=0,1$, presented in the top two rows of 
Fig.~\ref{fig:ttsa}.  Equations~\ref{eq:cT0I} and \ref{eq:cT1I} show that these  amplitudes are sensitive to the GPD $E$ and hence to the total angular momenta of quarks. 
These amplitudes are found to have substantial magnitudes with opposite signs but little kinematic dependence, possibly increasing in magnitude with $-t$.
Their opposite signs are expected from Eq.~\ref{eq:cT0I}.
%arise from the interplay\footnote{See footnote on previous page} between the coefficient $c_{1,\rmTP}^\rmI$ and the squared BH amplitude $\left|\rmT_{\rmBH}\right|^2$.
Also of interest is the amplitude $\AUTDVCS^{\sin(\phi-\phi_S)}$, shown in the top row of Fig.~\ref{fig:ttsa},
which Eq.~\ref{eq:cT0DVCS} suggests is also sensitive to the GPD $E$.
The overall result is non-zero by 2.8 times the total uncertainty. 
%Furthermore, as is shown in Eq.~\protect\ref{eq:cT0DVCS} this amplitude is expected to have a dependence on $Q^2$ that is different from that of the interference term.
These data tend to increase in magnitude at larger values of $Q^2$.
%\footnote{Dietmar: to check whether higher twist can appear in Eq.~\protect\ref{eq:cT0DVCS}. Not relevand anymore - the data INCREASES in magnitude with Q2, so certainly not twist-3.}
(In fixed-target experiments, $x_B$ and $Q^2$ are strongly correlated.)  
They provide the first experimental evidence for an azimuthal harmonic in the squared DVCS amplitude, in this case related to transverse target polarisation. 

The amplitude $A_{UT,I}^{\cos(\phi-\phi_S) \sin \phi}$ shown in the bottom row of Fig.~\ref{fig:ttsa} is sensitive mainly to the GPDs $\widetilde{H}$ and $\widetilde{E}$, while the contribution from the GPD $E$ is suppressed by an additional factor of $x_B$ (see Eq.~\ref{eq:sT1I}).
%The latter can also be seen from the model calculations shown in the same figure, where variations of the parameter settings for the GPD $E$ become manifest only at large values of $x_B$ or correspondingly large $Q^2$. In the model description, 
%The forward limit of the GPD $\widetilde{H}$ is fixed by the quark helicity distributions $\Delta q(x_B,Q^2)$, while the GPD $\widetilde{E}$ is expected to be dominated by the pion pole, which provides only a real part. 
The measured asymmetry amplitudes are consistent with zero.

The amplitudes represented by the unfilled symbols are expected to be suppressed, and are indeed found to be typically small.
However, values that depart from zero by more than twice the total uncertainty are found for the entire experimental acceptance for two of the four amplitudes in Fig.~\ref{fig:ttsa2} that receive a contribution from gluon helicity-flip, which are $\AUTDVCS^{\cos(\phi-\phi_S)\sin(2\phi)}$ and $\AUTI^{\cos(\phi-\phi_S)\sin(3\phi)}$. 
The asymmetry amplitudes related to the squared DVCS amplitude in the bottom two rows of Fig.~\ref{fig:ttsa2} correspond to coefficients that do not appear in Eq.~\ref{eq:DVCS2} as a consequence of the one-photon exchange approximation.
They are found to be consistent with zero.

\section{Comparison with theory}

The data are compared with various theoretical calculations to LO in 
$\alpha_{em}$ and $\alpha_s$, which do not account for the contributions 
of associated production. The calculations 
shown in Figs.~\ref{fig:bca}--\ref{fig:ttsa2} 
employ variants of a GPD model developed in Refs.~\cite {Van99,Goe01}. 
These are based on the widely used framework of double distributions
involving a product of PDFs representing the forward limit and a 
profile function representing the skewness dependence~\cite{Rad99}.
The forward limits of the GPDs $H_q$ are constructed using the MRST98~\cite{Mar98} parameterisation of PDFs evaluated at the measured $Q^2$ value for each data point.
More modern parameterisations are expected to result in a negligible
difference~\cite{Elli05}.
In the model description for $\widetilde H$ and $\widetilde E$, the
forward limit of the GPD $\widetilde H$ is fixed by the quark helicity
distributions $\Delta q(x_B,Q^2)$, while the GPD $\widetilde E$ is
evaluated from the pion pole, which  provides only the real part.
The 'profile parameters' $b_{val}$ and $b_{sea}$ control the skewness dependence
of GPDs for the valence and sea quarks, respectively~\cite{Rad99,Rad00}.  
The $t$ dependence of the GPDs is calculated in either the simplest ansatz where 
the $t$ dependence factorises from the $t$--independent part $H_q(x,\xi)$,
or in the Regge-inspired ansatz.
%\footnote{This seems to beg the question: how is this $t$--independent part
%distinguished in the Regge approach?}
The GPDs $H$ and $E$ are optionally modified by
the so-called ($t$--independent)
D--term~\cite{Pol99} contribution to the double-distribution
part of the GPD, with a value
calculated in the chiral--quark soliton model~\cite{Pet98}.
%In addition,
The twist-three GPDs are treated in the Wandzura-Wilczek approximation, and the gluon helicity-flip GPDs are not included.
The quark total angular momenta $J_q$ of quarks and antiquarks of flavour $q$ ($q=u$,$d$) enter as model parameters for the GPD $E$.
The strange sea is neglected. 
The computational program of Ref.~\cite{Van01} is used. 
The calculation is done at the average kinematics  of every bin (see Table~\ref{table:ttsa}).
For the comparison of the BCA amplitudes to the double-distribution  
model  shown in Fig.~\ref{fig:bca}, 
the model variations of interest are those that change the GPD $H$,
since the impact of the GPDs $\widetilde H$ and $E$ is suppressed at HERMES kinematic conditions
(see Eq.~\ref{eq:cU1I}).
Four different variants are selected by choosing
either a factorised (Fac) or a Regge-inspired (Reg) $t$ dependence, each with 
or without 
the contribution of the D-term.
While these four variants lead to very different model predictions for
$\AC^{\cos\phi}$ as illustrated in the figure, the variation of the profile parameters $b_{val}$ and $b_{sea}$
lead to smaller changes~\cite{Ell04}.
However, by comparing the data for the $\cos \phi$ amplitude with predictions of 
all four variants of this model in combination with four 
specific sets of values for the profile parameters, it is found that 
the calculation using the Regge-inspired $t$ dependence without the D--term and $b_{val}=\infty$,
$b_{sea} =1$ results in a confidence level much higher than all the alternatives.
Here, $b=1$ corresponds to a substantial skewness dependence, which is eliminated for $b=\infty$.
% preferred model  always excludes the contribution from the D-Term,
%and has either the one with the Regge-inspired $t$ dependence
%($b_{val} = b_{sea} = 1$ or $b_{val} = \infty$, $b_{sea} = 1$) or the factorised
%one ($b_{val} = 1$, $b_{sea} = \infty$ or $b_{val} = b_{sea} = \infty$).
%The calculation using the Regge-inspired $t$ dependence without the D-term and $b_{val}=\infty$,
%$b_{sea} =1$ results in the best confidence level when compared to the data. 
%Therefore all 
All the variants shown in Fig.~\ref{fig:bca} are calculated using
profile parameter values that yield the best agreement with data (see Table~\ref{table:bprof}).
%%\footnote{Insert a table with chi2 values (x,t,Q2) of the best variants (Eduard, Frank E.), 
%%take from Zhenyu's email from July 28. (modify fig.4 accordingly)
%% Eduard: modify the fig.4 so that each curve includes ONLY the version of b_val and b_sea which
%% best describes the data. For that, take the chi2 from Zhenyu's email (27-28.July), add 
%% together the x,t,Q2 and take the best one for figure AND table.
%% chi2 values for BCA, Ju=0.4, bv=bs=1
%% Fac, D:  41.6
%% FacnoD:  38.3
%% Reg, D: 210.7
%% RegnoD:  25.8
%% chi2 values for BCA, Ju=0.4, bv=1 bs=inf
%% Fac, D: 155.4
%% FacnoD:  22.1 <----
%% Reg, D: 355.3
%% RegnoD:  70.0
%% chi2 values for BCA, Ju=0.4, bv=inf bs=1
%% Fac, D:  23.4 <---- 
%% FacnoD:  67.5
%% Reg, D: 157.3 <----
%% RegnoD:  11.7 <----
%% chi2 values for BCA, Ju=0.4, bv=inf bs=inf
%% Fac, D: 103.9
%% FacnoD:  29.0
%% Reg, D: 279.1
%% RegnoD:  38.7
%% }
For any choice of the profile parameters,
the variant Regge with a D-term is excluded, while 
the factorised ansatz is disfavoured either with or without the D-term.
The factorised $t$ dependence is also disfavoured on 
theoretical grounds \cite{Gockeler:2006ui,Hagler:2007xi}.

The theoretical calculations of the TTSA amplitudes shown in
Figs.~\ref{fig:ttsa} and \ref{fig:ttsa2} are made using the
Regge inspired $t$ dependence without the D-term and 
$b_{val}=\infty$ and $b_{sea} =1$, a combination that is favoured 
by the BCA data as described above. 
However, while the calculated TTSA amplitudes are less sensitive to that
choice, some of them are quite sensitive to the choice of
the quark total angular momenta $J_q$~\cite{Elli05}.
This sensitivity is illustrated by 
the curves in Fig.~\ref{fig:ttsa} evaluated with three different values of 
$J_u$ (0.2, 0.4, 0.6), while fixing $J_d=0$~\cite{Elli05}. 
Although this model fails to describe the data for $A_{UT,\rmDVCS}^{\sin(\phi-\phi_S)}$ , the model curves confirm the expectation from 
Eqs.~\ref{eq:cT0DVCS}, \ref{eq:cT1I} and \ref{eq:cT0I} that 
the TTSA amplitudes $A_{UT,\rmDVCS}^{\sin(\phi-\phi_S)}$, $A_{UT,\rmI}^{\sin(\phi-\phi_S)\cos\phi}$ and $A_{UT,\rmI}^{\sin(\phi-\phi_S)}$ have significant sensitivity to $J_u$.
However, for this double-distribution model, 
the amplitudes related to the interference term
are in reasonable agreement with the data, of which $A_{UT,\rmI}^{\sin(\phi-\phi_S)\cos\phi}$ 
has the greatest sensitivity.
The curves in Fig.~\ref{fig:bca} and \ref{fig:ttsa2} are evaluated with fixed $J_u=0.4$ and $J_d=0$, 
since the sensitivity here to $J_u$ and $J_d$ is negligible.

\begin{figure}
\includegraphics[width=1\columnwidth]{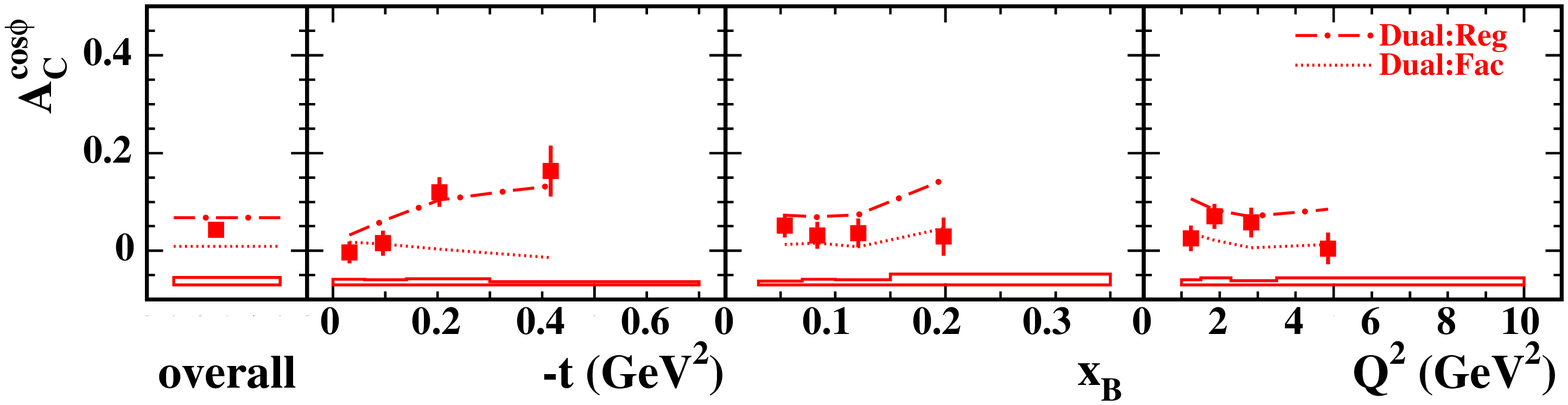}
\caption{Similar to the top row of Fig.~\ref{fig:bca}, except that the curves are 
  calculations~\cite{Vad07} based on the dual-parameterisation GPD model~\cite{Vad06}. 
See text for details.
\label{fig:bca-dual}}

\vspace{0.2cm}

\includegraphics[width=1\columnwidth]{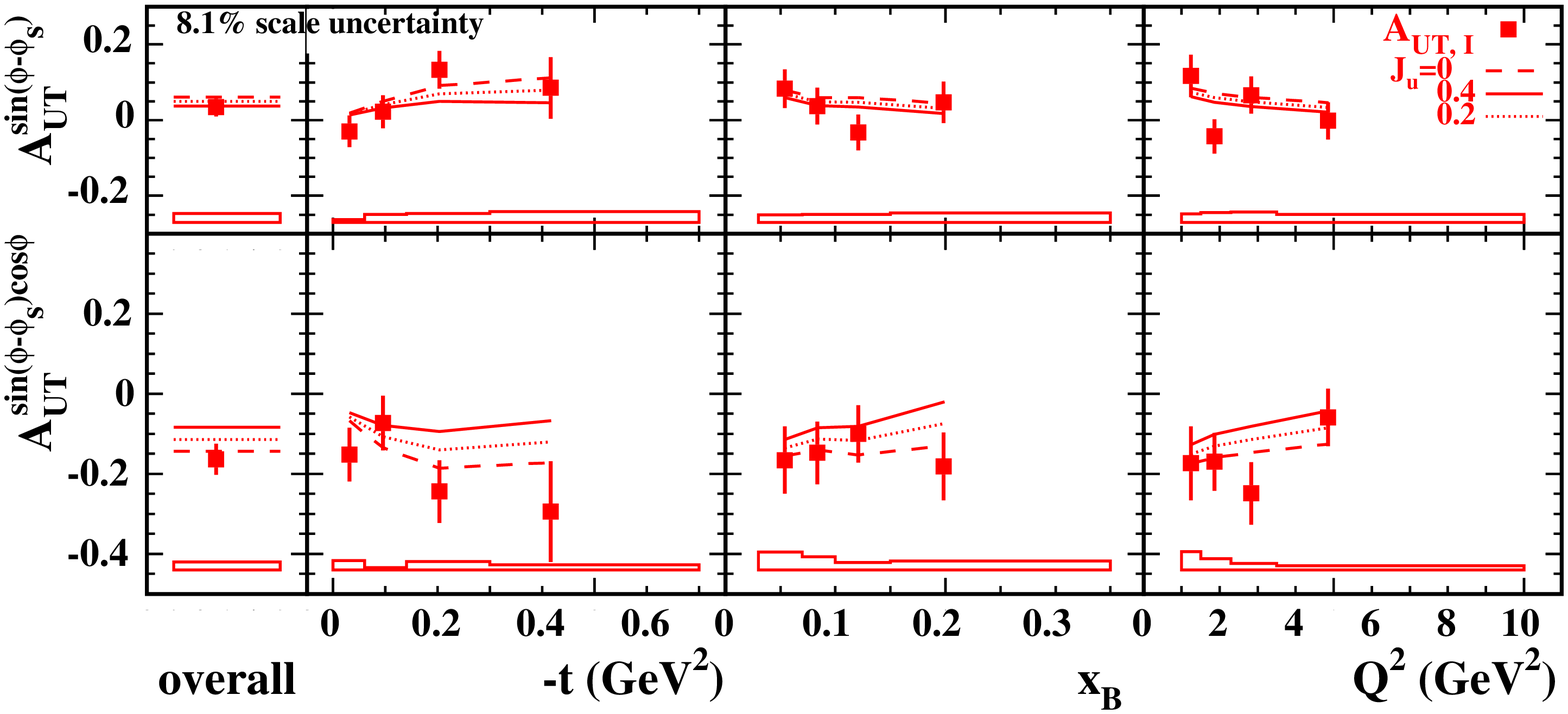}
\caption{Similar to top two rows of Fig.~\ref{fig:ttsa}, except that the curves are the
  calculations~\cite{Vad07} using the Regge-inspired form of the 
$t$ dependence in 
the dual-parameterisation GPD model~\cite{Vad06} 
(shown in Fig.~\ref{fig:bca-dual} as a dash-dotted curve), with three different 
values for the $u$-quark total angular momentum $J_u$ and fixed $d$-quark total angular 
momentum $J_d=0$. 
See text for details.
}
\label{fig:ttsa-dual}
\end{figure}

The BCA amplitude $\AC^{\cos\phi}$  and the TTSA amplitudes $A_{UT,\rmI}^{\sin(\phi-\phi_S)}$ and $A_{UT,\rmI}^{\sin(\phi-\phi_S)\cos\phi}$ are also compared to calculations
based on the `dual-parameterisation' model of GPDs~\cite{Pol02,Vad06} in Figs.~\ref{fig:bca-dual} and ~\ref{fig:ttsa-dual}.
Calculations for all other amplitudes are not shown since the model 
contains neither GPDs $\widetilde H$ and $\widetilde E$ nor higher-twist contributions.
For the BCA amplitude, the 
curves in Fig.~\ref{fig:bca-dual} are evaluated with fixed $J_u=J_d=0$, which best 
describes the TTSA amplitudes as discussed below.
The $t$ dependence of the GPDs is assumed to be either factorised and
exponential, or non-factorised and Regge-inspired~\cite{Goe01}.
Both choices describe the BCA data equally well, but with a smaller
confidence level ($\chi^2$/d.o.f.=2.2 for both Fac and Reg) than the 
favoured double-distribution model described above.
On the other hand, the existing beam-spin asymmetry 
data~\cite{Frank_proc} are better described by this dual-parameterisation model.
The Regge-inspired variant of this model with $J_u = 0$,
0.2, and 0.4, and $J_d=0$  is used for the calculations shown in Fig.~\ref{fig:ttsa-dual}. 
It is apparent that also in the dual-parameterisation model,
the amplitudes $A_{UT,\rmI}^{\sin(\phi-\phi_S)\cos\phi}$ and $A_{UT,\rmI}^{\sin(\phi-\phi_S)}$ are sensitive to $J_u$.

\section{Quark total angular momentum}

\begin{figure}
\begin{center}
\includegraphics[width=0.7\columnwidth]{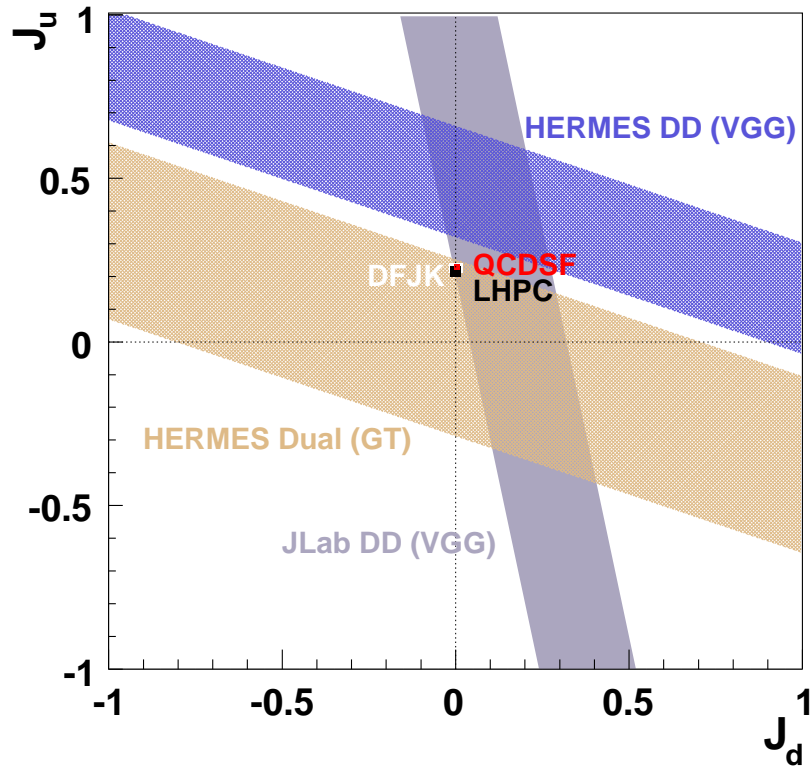}
\end{center}
\vspace*{-7mm}
\caption{Model-dependent constraints on $u$-quark total angular momentum $J_u$ vs $d$-quark total angular momentum $J_d$, obtained by comparing DVCS experimental results and theoretical calculations.
The  constraints  based on the HERMES data for the TTSA amplitudes  $A_{\rmTP}^{\sin{(\phi-\phi_S)}\cos{\phi}}$ and $A_{\rmTP,\rmI}^{\sin(\phi-\phi_S)}$ use the double-distribution (HERMES DD)~\cite{Van99,Goe01} or dual-parameterisation (HERMES Dual)~\cite{Vad06}  GPD models.
The additional band (JLab DD) is derived  from the comparison
of the double-distribution GPD model with neutron cross section data~\cite{JLABneutron}.
Also shown as small (overlapping) rectangles are results
% (statistical uncertainties only)
from lattice gauge theory by the 
QCDSF~\cite{QCDSF} and LHPC~\cite{Hagler:2007xi} collaborations, 
as well as a result for only the valence quark
contribution (DFJK) based on zero-skewness GPDs 
extracted from nucleon form factor data~\cite{dfjk,Kroll:2007wn}. 
The sizes of the small rectangles represent the 
statistical uncertainties of the lattice gauge results, 
and the parameter range for which a good DFJK fit to the 
nucleon form factor data was achieved. Theoretical uncertainties are unavailable.}  
\label{fig:constraint1}
\end{figure}

In either GPD model discussed above, 
the GPD $E$ is parameterised in terms of $J_u$ and $J_d$~\cite{Goe01,Elli05}. 
In both cases these two parameters
are fit to the measured overall 
TTSA amplitudes $A_{\rmTP,\rmI}^{\sin(\phi-\phi_S)\cos\phi}$ and
$A_{\rmTP,\rmI}^{\sin(\phi-\phi_S)}$ 
appearing in the left
column of Figs.~\ref{fig:ttsa} and~\ref{fig:ttsa-dual}, respectively.
The other parameters for the respective GPD models are the same as used
for the curves in Figs.~\ref{fig:ttsa} or \ref{fig:ttsa-dual}.
The area in the ($J_u$, $J_d$)-plane in which the reduced 
$\chi^2-\chi^2_{min}$ value is not larger than unity corresponds 
to a one-standard-deviation constraint on $J_u$ vs $J_d$.
This area is found to be one of the 
sloped bands in Fig.~\ref{fig:constraint1}, which in units of $\hbar$ can be represented 
in the case of the double-distribution model as
\begin{equation}
J_u+J_d/2.8=0.49\pm0.17(\mathrm{exp_{tot}}),\label{eq:constraint}
\end{equation}
and in the case of the dual-parameterisation model as
\begin{equation}
J_u+J_d/2.8=-0.02\pm0.27(\mathrm{exp_{tot}}).\label{eq:constraint-dual}
\end{equation}
The uncertainty is propagated from the total experimental uncertainty in 
the measured TTSA amplitudes. 
This uncertainty dominates the effects of variations within either of 
the GPD models of the values of profile parameters $b$ and the inclusion 
of the D term.
The variation of the value of $J_u+J_d/2.8$ obtained by  fitting the amplitudes in the bins of the three kinematic variables is found to be not larger than $\pm 0.15$. 
The results from fitting the two azimuthal 
amplitudes separately were found to be consistent.
The large difference between the constraints obtained using the double-distribution
and dual-parameterisation models is an indication of a large model dependence of 
the $(J_u,J_d)$ constraint obtained, which may be
related to the failure of both models to fully describe all
other available DVCS data.
Both constraints are consistent with results, also shown in Fig.~\ref{fig:constraint1}, from
unquenched lattice gauge simulations by the QCDSF~\cite{QCDSF} and the LHPC~\cite{Hagler:2007xi}
collaborations.
The statistical uncertainties of
the lattice gauge results are comparable to the size of the
plotted symbols.
The QCDSF calculation of the first moments of the GPDs, the so-called
generalized form factors, is based on a simulation using dynamical Wilson
fermions with pion masses down to 350 MeV.
The dynamical LHPC calculation is based on a hybrid approach of rooted
staggered sea and domain wall valence quarks.
In both calculations the generalized form factors have been simultaneously
extrapolated in $t$ and
$m_\pi^2$ to $t=0$ and the physical point, respectively, using the same results from
chiral perturbation theory.
Both calculations include only contributions from connected diagrams.
The uncertainties are primarily statistical but include some systematic
uncertainties from the chiral, continuum and infinite volume extrapolations.
% No estimates are available yet for theoretical systematic uncertainties.
Also shown in Fig.~\ref{fig:constraint1} is a result
for only the valence contribution to the quark total angular momenta~\cite{dfjk,Kroll:2007wn}.
It is based on the extraction of zero-skewness GPDs from nucleon form
factor data, assuming handbag diagram dominance and exploiting well known sum rules.
The size of the plotted symbol corresponds to the parameter range for which 
a good fit to the 
nucleon form factor data was achieved.

\begin{table}[h]\scriptsize
\caption{\label{table:ttsa} Results of particular interest for the asymmetry amplitudes of the asymmetries with respect to the beam charge and transverse target polarisation for the exclusive sample.}
\begin{tabular}{c|c|c|c|c|r|r|r}
\br
 \multicolumn{2}{c|}{kinematic bin} & $\left<-t\right>$ & $\left<x_B\right>$ & $\left<Q^2\right>$ & $\AC^{\cos{(0\phi)}}$\hspace{1cm} & $\AC^{\cos{\phi}}$\hspace{1.4cm} & $\mathrm{A_{\rmTP,\,\rmDVCS}^{\sin{(\phi-\phi_S)}}}$\hspace{0.8cm}  \\
  \multicolumn{2}{c|}{}       & (GeV$^2$) & & (GeV$^2$) & $\mathrm{\pm\delta_{stat}\pm\delta_{syst}}$ & $\mathrm{\pm\delta_{stat}\pm\delta_{syst}}$ & $\mathrm{\pm\delta_{stat}\pm\delta_{syst}}$  \\
\mr
  \multicolumn{2}{c|}{overall} & 0.12 & 0.09 & 2.5 & $-0.011\pm0.010\pm0.017$ & $0.043\pm0.014\pm0.015$ & $-0.073\pm0.024\pm0.008$ \\
\mr
          & 0.00--0.06 & 0.03 & 0.08 & 1.9 & $0.010\pm0.016\pm0.010$ & 
$-0.003\pm0.022\pm0.012$ & $-0.070\pm0.041\pm0.009$											 \\
\multirow{4}{2mm}{\begin{sideways}\hspace{-0.25cm}\parbox{3mm}{$-t$(GeV$^2$)}\end{sideways}}      & 0.06--0.14 & 0.10 & 0.10 & 2.5 & $-0.006\pm0.019\pm0.017$ & $0.015\pm0.026\pm0.011$ & $-0.067\pm0.043\pm0.017$											 \\
 & 0.14--0.30 & 0.20  & 0.11  & 2.9 & $-0.026\pm0.022\pm0.018$ & $0.120\pm0.030\pm0.012$ & $-0.066\pm0.050\pm0.011$											 \\
          & 0.30--0.70 & 0.42  & 0.12  & 3.5 & $-0.074\pm0.036\pm0.024$ & $0.163\pm0.052\pm0.007$ & $-0.153\pm0.080\pm0.015$											 \\
\mr
          & 0.03--0.07 & 0.10 & 0.05 & 1.5 & $-0.006\pm0.017\pm0.009$ & $0.051\pm0.024\pm0.008$ & $-0.008\pm0.051\pm0.008$											 \\
\multirow{4}{2mm}{\begin{sideways}\parbox{2mm}{$\quad x_B$}\end{sideways}}           & 0.07--0.10 & 0.10 & 0.08 & 2.2  & $-0.027\pm0.019\pm0.014$ & $0.032\pm0.027\pm0.012$ & $-0.079\pm0.049\pm0.010$											 \\ 
   & 0.10--0.15 & 0.13 & 0.12 & 3.1   & $0.000\pm0.022\pm0.014$  & $0.037\pm0.030\pm0.011$ & $-0.105\pm0.047\pm0.013$											 \\
          & 0.15--0.35 & 0.20 & 0.20 & 5.0   & $-0.003\pm0.029\pm0.021$ & $0.029\pm0.039\pm0.022$ & $-0.201\pm0.058\pm0.027$											 \\
\mr
          & 1.0--1.5 & 0.08 & 0.06 & 1.2   & $-0.014\pm0.019\pm0.016$ & $0.025\pm0.026\pm0.011$ & $0.044\pm0.056\pm0.012$ 											 \\
\multirow{4}{2mm}{\begin{sideways}\hspace{-0.25cm}\parbox{2mm}{$Q^2$(GeV$^2$)}\end{sideways}}     & 1.5--2.3 & 0.10 & 0.08 & 1.9    & $-0.004\pm0.018\pm0.016$ & $0.070\pm0.026\pm0.015$ & $-0.080\pm0.046\pm0.010$											 \\
 & 2.3--3.5 & 0.13 & 0.11 & 2.8     & $-0.023\pm0.021\pm0.015$ & $0.058\pm0.030\pm0.008$ & $-0.113\pm0.049\pm0.012$											 \\
          & 3.5--10.0 & 0.19 & 0.17 & 4.9    & $-0.003\pm0.023\pm0.016$ & $0.005\pm0.032\pm0.014$ & $-0.143\pm0.048\pm0.015$											 \\
\br
\multicolumn{8}{c}{}\\
\br
 \multicolumn{2}{c|}{kinematic bin} & $\left<-t\right>$ & $\left<x_B\right>$ & $\left<Q^2\right>$ & $\mathrm{A_{\rmTP,\,\rmI}^{\sin{(\phi-\phi_S)}}}$\hspace{0.8cm} & $\mathrm{A_{\rmTP,\,\rmI}^{\sin{(\phi-\phi_S)}\cos{\phi}}}$\hspace{0.3cm} & $\mathrm{A_{\rmTP,\,\rmI}^{\cos{(\phi-\phi_S)}\sin{\phi}}}$\hspace{0.3cm} \\
  \multicolumn{2}{c|}{}       & (GeV$^2$) & & (GeV$^2$) & $\mathrm{\pm\delta_{stat}\pm\delta_{syst}}$ & $\mathrm{\pm\delta_{stat}\pm\delta_{syst}}$ & $\mathrm{\pm\delta_{stat}\pm\delta_{syst}}$ \\
\mr
  \multicolumn{2}{c|}{overall} & 0.12 & 0.09 & 2.5 & $0.035\pm0.024\pm0.024$ & $-0.164\pm0.039\pm0.023$ & $0.005\pm0.040\pm0.015$ \\
\mr
          & 0.00--0.06 & 0.03 & 0.08 & 1.9 &  $-0.030\pm0.031\pm0.008$ & $-0.152\pm0.068\pm0.026$ & $-0.100\pm0.069\pm0.044$ \\
\multirow{4}{2mm}{\begin{sideways}\hspace{-0.27cm}\parbox{2mm}{$-t$(GeV$^2$)}\end{sideways}}     
	  & 0.06--0.14 & 0.10 & 0.10 & 2.5 &  $0.022\pm0.044\pm0.021$  & $-0.073\pm0.068\pm0.008$ & $0.054\pm0.076\pm0.030$ \\
	  & 0.14--0.30 & 0.20  & 0.11  & 2.9 &  $0.133\pm0.050\pm0.025$  & $-0.244\pm0.078\pm0.028$ & $0.144\pm0.083\pm0.020$ \\
          & 0.30--0.70 & 0.42  & 0.12  & 3.5 &  $0.085\pm0.082\pm0.028$  & $-0.294\pm0.126\pm0.026$ & $0.024\pm0.113\pm0.029$ \\
\mr
          & 0.03--0.07 & 0.10 & 0.05 & 1.5 &  $0.083\pm0.051\pm0.021$  & $-0.166\pm0.084\pm0.047$ & $-0.034\pm0.081\pm0.025$ \\
\multirow{4}{2mm}{\begin{sideways}\parbox{2mm}{$\quad x_B$}\end{sideways}} 
          & 0.07--0.10 & 0.10 & 0.08 & 2.2  &  $0.037\pm0.048\pm0.021$  & $-0.148\pm0.078\pm0.034$ & $-0.078\pm0.080\pm0.015$ \\
	  & 0.10--0.15 & 0.13 & 0.12 & 3.1   &  $-0.033\pm0.048\pm0.021$ & $-0.100\pm0.072\pm0.020$ & $0.078\pm0.073\pm0.025$ \\
          & 0.15--0.35 & 0.20 & 0.20 & 5.0   &  $0.048\pm0.055\pm0.024$  & $-0.182\pm0.084\pm0.026$ & $0.066\pm0.088\pm0.056$ \\
\mr
          & 1.0--1.5 & 0.08 & 0.06 & 1.2   &  $0.117\pm0.056\pm0.024$  & $-0.174\pm0.092\pm0.047$ & $-0.034\pm0.093\pm0.018$ \\
\multirow{4}{2mm}{\begin{sideways}\hspace{-0.27cm}\parbox{2mm}{$Q^2$(GeV$^2$)}\end{sideways}}
          & 1.5--2.3 & 0.10 & 0.08 & 1.9    &  $-0.043\pm0.046\pm0.026$ & $-0.170\pm0.073\pm0.031$ & $-0.036\pm0.079\pm0.020$ \\
	  & 2.3--3.5 & 0.13 & 0.11 & 2.8     &  $0.066\pm0.049\pm0.028$  & $-0.249\pm0.078\pm0.025$ & $0.028\pm0.076\pm0.026$ \\
          & 3.5--10.0 & 0.19 & 0.17 & 4.9    &  $-0.002\pm0.049\pm0.020$ & $-0.059\pm0.072\pm0.011$ & $0.056\pm0.079\pm0.035$ \\
\br
\end{tabular}
% \vspace*{0.5cm}
%\end{indented}
\end{table}

\section{Conclusions}

Transverse target-spin azimuthal asymmetries in electroproduction of real photons are measured for the first time, and for both beam charges. 
A combined fit of this data set separates for the first time the azimuthal 
harmonics of the squared DVCS amplitude and the interference term. 
The extracted charge asymmetry of the interference term is much 
more precise than 
previously published results, and  constrains models 
for Generalized Parton Distributions.
By comparing GPD-model calculations with extracted azimuthal asymmetry amplitudes associated with both beam charge and transverse-target polarisation, a model-dependent constraint on the total angular momenta carried by $u$ and $d$-quarks in the nucleon is obtained as 
$J_u+J_d/2.8=0.49\pm0.17(\mathrm{exp_{tot}})$ using a double-distribution GPD model, and
$J_u+J_d/2.8=-0.02\pm0.27(\mathrm{exp_{tot}})$ using the dual-parameterisation model. 
Thus, such data have the potential to provide quantitative information 
about the spin content of the nucleon when GPD models become available that
fully describe all existing DVCS data.

\begin{table}[h]%\scriptsize
\vspace*{-4mm}
\caption{Systematic uncertainties of the results of particular interest for the 
azimuthal amplitudes of the asymmetries with respect to the beam charge and transverse 
target polarisation for the exclusive sample. 
Those results involving transverse target polarisation are also subject to an additional 8.1\% scale uncertainty from the determination of the target polarisation.\label{table:syst}}
\begin{indented}
\item[]\vspace*{4mm}\begin{tabular}{c|c|c|c|c|c|c}
\br
Source & \begin{sideways}$\AC^{\cos{(0\phi)}}$\end{sideways} & \begin{sideways}$\AC^{\cos{\phi}}$\end{sideways} & \begin{sideways}$A_{\rmTP,\,\rmDVCS}^{\sin{(\phi-\phi_S)}}$\end{sideways} & \begin{sideways}$A_{\rmTP,\,\rmI}^{\sin{(\phi-\phi_S)}}$\end{sideways} & \begin{sideways}$A_{\rmTP,\,\rmI}^{\sin{(\phi-\phi_S)}\cos{\phi}}$\end{sideways} & \begin{sideways}$A_{\rmTP,\,\rmI}^{\cos{(\phi-\phi_S)}\sin{\phi}}$\end{sideways} \\ 
\br
$M_X^2$ shift             & 0.004 & 0.001 & 0.000 & 0.000 & 0.001 & 0.001 \\
% Acceptance and bin width  & 0.016 & 0.018 & 0.003 & 0.021 & 0.021 & 0.013 \\
Background correction     & 0.000 & 0.001 & 0.005 & 0.001 & 0.004 & 0.000 \\
Calorimeter calibration      & 0.001 & 0.003 & 0.002 & 0.002 & 0.005 & 0.003 \\
Acceptance, bin width, alignment
& 0.017 & 0.015 & 0.002 & 0.024 & 0.019 & 0.014 \\
\br
\end{tabular}
\end{indented}
% \vspace*{0.5cm}
\end{table}

\begin{table}[h]
\vspace*{-3mm}
\caption{The quality of agreement between the measured $A_C^{\cos(\phi)}$
asymmetry amplitude and four variants of the double-distribution model. 
These $\chi^2$ values are based on sums over the 12 bins
in $-t$, $x_B$ and $Q^2$, without considering the
correlations among them because of their sharing of
some events.  The values given are the sums divided
by 12. Values shown
in bold face denote for each variant the profile parameter set that
yields the best agreement with the data, chosen to produce 
the curves in Fig.~\ref{fig:bca}.}
\label{table:bprof}
\begin{indented}
\item[]\vspace*{3mm}
\begin{tabular}{l|r|r|r|r}
\br
$\chi^2$/d.o.f. ($J_u=0.4$) & $b_v=b_s=1$ & $b_v=1$, $b_s=\infty$ &
	$b_v=\infty$, $b_s=1$ & $b_v=b_s=\infty$ \\
\br
Fac, D     &  4.2 & 16.0 & \textbf{2.3} & 10.5 \\
Fac, no D  &  4.5 & \textbf{2.3} &  7.9 &  3.3 \\ 
Reg, D     & 22.2 & 37.4 & \textbf{16.5}& 29.2 \\ 
Reg, no D  &  2.5 & 7.0  & \textbf{1.1} &  3.8 \\

\br
\end{tabular}
\end{indented}
\end{table}

\ack
We gratefully acknowledge the DESY management for its support and the staff at DESY and the collaborating institutions for their significant effort.  We are also grateful for careful reading and advice
by Markus Diehl and Dieter M{\"u}ller.
This work was supported by the FWO-Flanders, Belgium;
the Natural Sciences and Engineering Research Council of Canada;
the National Natural Science Foundation of China;
the Alexander von Humboldt Stiftung;
the German Bundesministerium f\"ur Bildung und Forschung (BMBF);
the Deutsche Forschungsgemeinschaft (DFG);
the Italian Istituto Nazionale di Fisica Nucleare (INFN);
the MEXT, JSPS, and COE21 of Japan;
the Dutch Foundation for Fundamenteel Onderzoek der Materie (FOM);
the U. K. Engineering and Physical Sciences Research Council, 
the Particle Physics and Astronomy Research Council and the Scottish Universities Physics Alliance;
the U. S. Department of Energy (DOE) and the National Science Foundation (NSF)
and the Ministry of Trade and Economical Development and the Ministry of Education and Science of Armenia.

%\newpage

%\begin{landscape}
%\end{landscape}

% -------------------- Bibliography (numbered style) --------------------
% \newpage

\end{document}